\documentclass{aa}

\usepackage{graphicx,natbib}
\bibpunct{(}{)}{;}{a}{}{,} % to follow the A&A style
\usepackage[varg]{txfonts}
\usepackage[colorlinks=true,allcolors=blue]{hyperref}
\usepackage[normalem]{ulem}
\usepackage{threeparttable}

\newcommand{\refereed}[1]{{#1}}

\begin{document}

%\title{How leaky?: A parameter space exploration of leaky dust traps to quantify the transport of dust and volatiles from outer to inner disc}

\title{How leaky? A large parameter study of leaky dust traps to quantify the transport of pebbles and ice in protoplanetary discs}

   \author{Adrien Houge\inst{1}
    \and
    Anders Johansen\inst{1,2}
    \and
    Andrea Banzatti\inst{3}
    \and
    Sierra Grant\inst{4}
      }

    \institute{Center for Star and Planet Formation, Globe Institute, University of Copenhagen, Øster Voldgade 5-7, DK-1350 Copenhagen, Denmark\\
    \email{adrien.houge@sund.ku.dk}
    \and
    Lund Observatory, Department of Physics, Lund University, Box 43, 221 00 Lund, Sweden
    \and
    Department of Physics, Texas State University, 749 North Comanche Street, San Marcos, TX 78666, USA
    \and
    Earth and Planets Laboratory, Carnegie Institution for Science, 5241 Broad Branch Road, NW, Washington, DC 20015, USA
    }    
    
    \date{Received ; }

%\author[0000-0001-8790-9011]{Adrien Houge}
%\author[0000-0002-5893-6165]{Anders Johansen}

% \abstract{}{}{}{}{} 
% 5 {} token are mandatory

    \abstract
    {In protoplanetary discs, the presence of dust traps can significantly alter the transport of solids from the outer to the inner regions, and hence they are often invoked as an explanation for the chemical diversity of inner discs observed with JWST (e.g., varying oxygen abundances and C/O ratios). As a detailed treatment of dust transport around dust traps is computationally expensive, earlier works investigating the impact of outer traps on the inner disc composition have often used simplified dust models representing the size distribution with a single effective size and drift speed. In this paper, we revisit the impact of outer traps on dust transport using the state-of-the-art one-dimensional dust evolution code \texttt{DustPy}, which simulates the transport and evolution of dust particles including detailed coagulation and fragmentation. We quantify and map the leakiness of dust traps across a broad parameter space, performing over 300 simulations while varying the disc viscosity, turbulence strength, planet mass and location, and dust fragmentation velocity. We find that dust traps are leakier than previously thought, on a broader parameter space, such that most outer traps ($r > 5 \mathrm{~au}$) will result in a long-lived oxygen-rich inner disc with gas-phase $\mathrm{C/O} < 1$. In similar conditions (e.g., carved by the same planet mass), we find inner traps are much leakier than outer traps, though their relative efficiency in reducing the pebble flux is time-dependent. Highly blocking traps altering the inner disc composition dramatically (leading, e.g., to $\mathrm{C/O} > 1$) are possible to set up but necessitate low viscosity ($\alpha_\mathrm{visc} \leq 10^{-3}$) and weak turbulence ($\delta_\mathrm{turb} \leq 10^{-4}$), along with efficient planetesimal formation by the streaming instability. In that case, we find that is the formation of planetesimals, rather than the dust traps themselves, that is capable of significantly altering the inner disc composition.}
    %{} {} {} {}


    \titlerunning{How leaky?}
    \authorrunning{Houge et al.}
    
   \maketitle

\section{Introduction} 
\label{sec:intro}

The characterisation of the warm inner ($r < 5 \mathrm{~au}$) region of protoplanetary discs from infrared observations offers a unique opportunity to characterise the chemical reservoir of discs and how the observed reservoir may connect to the transport of volatiles by icy pebbles from the outer regions of the disc \citep{banzatti2020hints, banzatti2023JWST}. First results obtained with the Spitzer Space Telescope revealed a significant diversity across molecular spectra of the inner part of protoplanetary discs \citep[e.g.,][]{carr2008organics, pontoppidan2010spitzer, salyk2011rovibrational, salyk2011spitzer, pontoppidan2014volatiles, banzatti2017depletion}, with for example the $\mathrm{H_2O}$/$\mathrm{HCN}$ ratio varying with stellar mass \citep{pascucci2009different, pascucci2013atomic} and disc mass \citep{najita2013hcn}. More recently, the James Webb Space Telescope (JWST) has allowed astronomers to acquire a wealth of infrared data with a much higher sensitivity and spectral resolving power, which confirmed the significant diversity in molecular spectra of inner discs \citep[e.g.,][]{perotti2023water, xie2023water, grant2023minds, grant2024minds, colmenares2024jwst, frediani2025xue, long2025first, grant2025minds, arulanantham2025jdisc, banzatti2025water}. Several works have developed models to explain the observed compositional diversity in discs, notably involving dust optical depth effects \citep{najita2011formation, woitke2018modelling, greenwood2019effects, sellek2025co2, houge2025smuggling}, planetesimal formation \citep{najita2013hcn, danti2023composition}, refractory carbon decomposition \citep{houge2025burned}, and variations in the disc evolution with stellar mass \citep{mah2023close, sellek2025chemical}. Still, one mechanism has largely been invoked to explain the observed chemical diversity, as it may be very common in discs \citep{andrews2018disk}: namely the trapping of dust in outer pressure bumps.

As a result of the sub-Keplerian structure of the gas in protoplanetary discs, millimetre to centimetre-sized particles, so-called pebbles, acquire a radial drift motion that is directed inwards, toward regions of higher pressure \citep{weidenschilling1997origin}. Nevertheless, in the presence of a local pressure maxima, such as those arising from the presence of a massive planet carving a gap, inward-drifting pebbles can become trapped. On top of locally increasing the dust density and facilitating planetesimal formation \citep{eriksson2020pebble, eriksson2021fate}, dust traps may prevent the delivery of icy solids to the warm inner disc, where their volatile content sublimates into the gas-phase \citep[e.g.,][]{cyr1998distribution, cuzzi2004material, ciesla2006evolution, krijt2016tracing, booth2017chemical, houge2025smuggling, williams2025locked, wang2025solving}. Depending on the position of the local pressure maximum relative to different icelines, this process controls how pebbles deliver different molecular species to the inner disc, hence resulting in a diversity of chemical inventories \citep{kalyaan2021linking, kalyaan2023effect, mah2024mind, sellek2025co2}. Interestingly, a similar mechanism is often invoked in the Solar Nebula to explain the isotopic dichotomy between carbonaceous (CC) and non-carbonaceous (NC) meteorites, based on the assumption that Jupiter may have formed early and carved a gap that separated the disc into two distinct reservoirs \citep{vankooten2016isotopic, kruijer2017age, alibert2018formation}.

The underlying assumption behind invoking local pressure maxima to explain the chemical diversity seen with JWST, or the NC/CC dichotomy in the Solar System, is that such traps must trap a significant fraction of the incoming pebble flux from the outer disc. However, there has been a growing amount of evidence that dust traps may be leaky, slowly losing mass and delivering material to the inner disc. Dust traps experience leakage because while pebbles are trapped, small grains remain well coupled to the gas and continue to move inward with the gas flow. This results in a filtering of particles based on particle size \citep[e.g.,][]{rice2006dust, zhu2012dust, liu2022natural, vanclepper20253D, facchini20262}, allowing for a part of the solid mass to leak through the dust trap and make it to the inner disc. While only a small fraction of the total solid mass is stored in small grains, considering the mass exchange between pebbles and small grains via collisional fragmentation dramatically increases the mass flux leaking through the trap  \citep{pinilla2012ring, pinilla2016tunnel, drazkowska2019including, stammler023leaky, pinilla2024survival, homma2024isotopic, tong2026turbulence}. Beyond theory, the leakiness of dust traps has also been suggested from observational measurements, such as the presence of an inner dust disc in PDS 70 \citep{perotti2023water, pinilla2024survival} and other transition discs \citep{espaillat2010unveiling, benisty2010complex, olofsson2013sculpting, pinilla2016tunnel}. In PDS 70, additional evidence of the trap leakiness comes from analysis of the trap width with ALMA Band 9 showing the radial diffusion of small grains \citep{sierra2025leaky} along with the inner disc dust composition that is consistent with the mixing of solids from the outer disc reservoir \citep{jang2024mineralogy}. Last, the detection and variation in cold water vapour emission measured with JWST in a variety of discs hosting deep or shallow dust traps resolved by ALMA hints that dust traps can be leaky, with varying leakiness depending on trap properties \citep{banzatti2023JWST, banzatti2025water, arulanantham2025jdisc, gasman2025minds}.

%As mentioned in Sect.~\ref{sec:intro}, a number of mechanisms have been suggested to explain this chemical diversity. For example, variations in the C/O ratio may be due to the impact of stellar mass and viscous evolution \citep{mah2023close, sellek2025chemical} or the processing of refractory carbon \citep{houge2025burned}. The absence of $\mathrm{H_2O}$ can be due to the formation of a cavity extending beyong the water iceline \citep{vlasblom2024mid}. 

%Yet, in this picture the impact of a pressure bump depends sensitively on its position. Because we do not resolve them, it can technically be adjusted to reproduce a wide range of chemical outcomes. Its ability to match the observed diversity makes it an appealing explanation, but does not constitute an evidence for the existence of fully blocking pressure bumps. Moreover, when pressure bumps are actually resolved from ALMA observations, no key differences in JWST spectra is observed \Adrien{Gasman}.

% mention M_peb in HL Tau, and Kama/McClure stuff

%However, this filtration process sensitively depends on the pressure gradient at the planet-induced gap outer edge and it only operates for large particles whose drift velocities are significant. Small particles strongly coupled to the gas flow can still penetrate the planet-induced gap to the inner disk. More importantly, dust diffusion may significantly reduce the filtration efficiency even for the big particles (Ward 2009).$

Nevertheless, earlier disc-wide models designed to study the interplay of dust evolution (coagulation, drift) and disc chemistry (sublimation, advection, diffusion) in the presence of substructures \citep{kalyaan2021linking, kalyaan2023effect, mah2024mind, sellek2025co2} found that dust traps can be quite efficient at blocking icy pebbles and controlling the volatile delivery to the inner disc, especially inner traps \citep{kalyaan2023effect, krijt2025cosmic}. In fact, for a variety of trap profiles, \citet{mah2024mind} found that the presence of an outer disc dust trap can result in a depletion of oxygen to the inner disc, due to the blocking of icy pebbles, and a gas-phase C/O ratio above unity. However, some of these earlier works focusing on the impact of traps on the inner disc composition used simpler dust evolution models such as the two-population algorithm \citep[\texttt{two-pop-py}, see][]{birnstiel2012simple}, which evolves the whole dust size distribution with two particle sizes (small grains and pebbles) moving at a common speed. Though the two-population algorithm has been proven effective in modelling smooth discs \citep{birnstiel2012simple}, it is not clear how well it performs for discs hosting dust traps as it does not include collisional fragmentation, an essential ingredient of the leakiness of dust traps \citep[e.g.,][]{stammler023leaky}.

To properly assess the role of dust traps in shaping (or not) the chemical diversity measured in the inner part of protoplanetary discs, it is thus required to revise our models of leaky traps with full grain size distribution models, and explore the role of different key parameters in shaping the leakiness. In this paper, we therefore perform a large-scale parameter space exploration of dust traps with \texttt{DustPy} \citep{stammler2022dustpy}, a full grain size distribution code self-consistently simulating the motion of dust of different sizes and the mass exchanges between them, via coagulation and fragmentation. Our goal is to map out how dust traps trap dust and pebbles based on key parameters, quantify leaking across a large variety of possible discs, and identify the different regimes. As such, we aim to show whether dust traps can meaningfully impact spectra obtained with JWST, and if so, in which area of the parameter space sufficient trapping happens.

The paper is organised as follows. In Section~\ref{sec:methods}, we detail our disc model and parameter space exploration with \texttt{DustPy}, along with how we impose our gap profile and calculate planetesimal formation. We present our results in Section~\ref{sec:results}, starting with a fiducial model to introduce relevant quantities and notions (Section~\ref{sec:results_structured_vs_smooth}), before diving into our complete parameter space study (Section~\ref{sec:results_paramspace}). We discuss the implications of our findings in Section~\ref{sec:discussion}, before giving our conclusion in Section~\ref{sec:conclusions}.

\section{Methods} 
\label{sec:methods}

\begin{figure*}
    \centering
    \includegraphics[width=\textwidth]{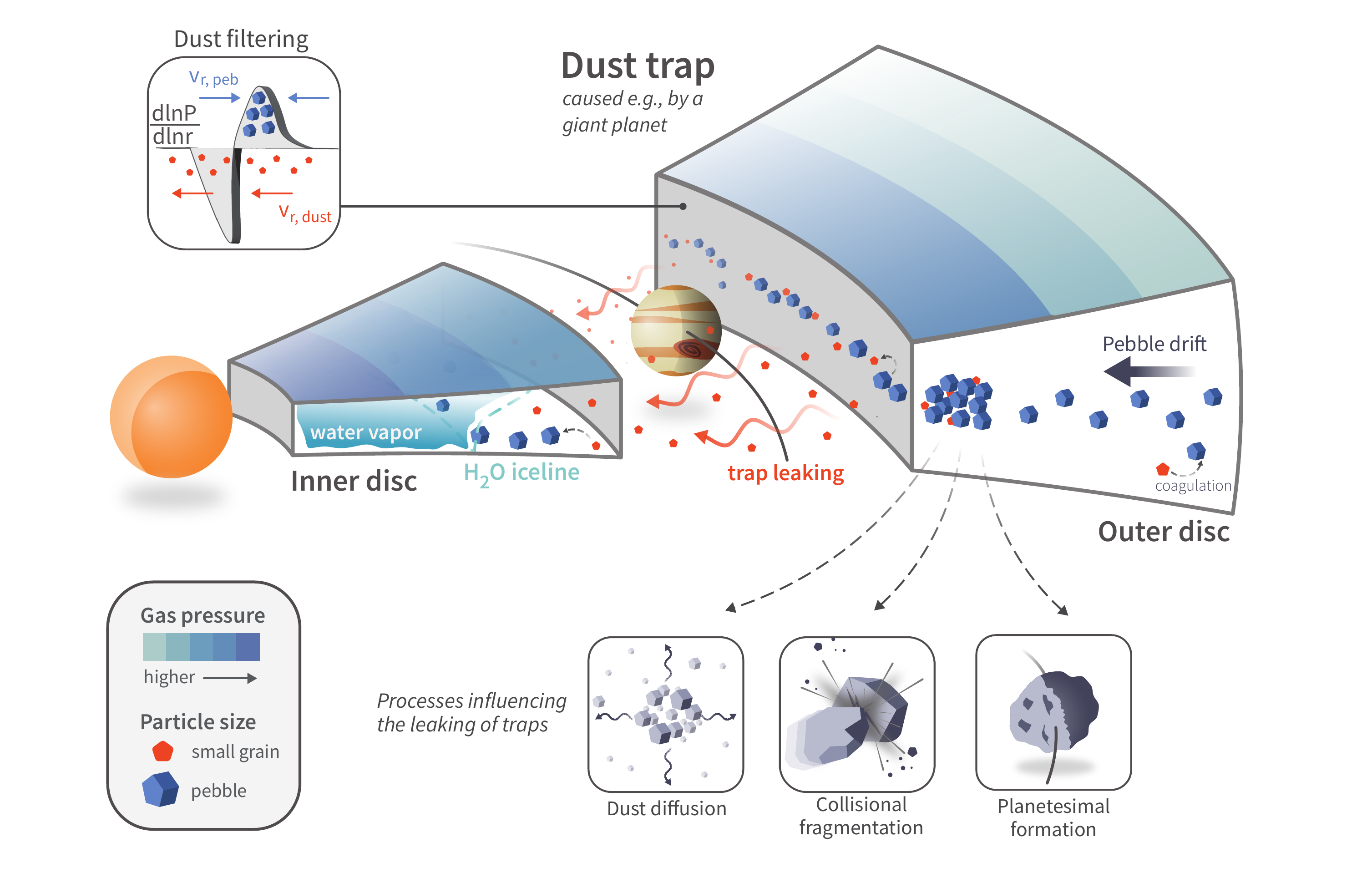}
    \caption{Schematic representing a protoplanetary disc with a gap created by the presence of a giant planet. The gas pressure profile is modified, resulting in the filtering of particles: pebbles are blocked while small grains leak past the gap. The key processes influencing the leakiness of dust traps included in our work are highlighted.}
    \label{fig:paper_main_schematic}
\end{figure*}

In this section, we describe our model to simulate the evolution of dust particles in a protoplanetary disc hosting a dust trap, along with a description of our parameter space study. We illustrate our approach in Fig.~\ref{fig:paper_main_schematic}.

\subsection{Dust and disc model}
\label{sec:methods_dust_disc_model}

To self-consistently simulate the evolution of dust particles in a protoplanetary disc with a local pressure maximum (i.e., a dust trap) and the potential leaking of small grains through it, it is necessary to use a state-of-the-art dust evolution code that includes coagulation, fragmentation and transport for a full dust size distribution \citep{drazkowska2019including}. We choose to use the dust evolution code \texttt{DustPy}\footnote{\refereed{We used \texttt{DustPy} v1.0.8 for the simulations presented in this paper.}} \citep{stammler2022dustpy}, often used by previous works\footnote{Note that as it is based on the Smoluchowski approach and discretized only in the space and size dimensions, \texttt{DustPy} cannot self-consistently include the evolution of the dust porosity.} to simulate dust evolution in planet-hosting discs \citep[e.g.,][]{pinilla2012ring, pinilla2016tunnel, stammler023leaky, pinilla2024survival}.

We adopt standard parameters to model the star and the disc, initializing the simulation with a disc of total mass $M_\mathrm{disc} = 0.05~\mathrm{M_\odot}$ (gas and dust) surrounding a solar-mass star with a stellar luminosity $1 \mathrm{~L}_\odot$. The disc temperature $T$ is governed by the stellar irradiation ($T \propto r^{-1/2}$) and we ignore the contribution of viscous heating, as it may be inefficient if disc winds dominate the angular momentum transport \citep{mori2019temperature, mori2021evolution}. We assume that the initial gas surface density profile follows a tapered power-law \citep[]{lynden1974evolution, hartmann1998accretion} with a characteristic radius $R_\mathrm{c}= 100 \mathrm{~au}$ and power-law $\gamma=1$. The initial dust-to-gas ratio is set to $\delta_\mathrm{d2g} = 0.01$. We initialize the simulation with a population of dust grains following an MRN size distribution \citep[$n(a) \propto a^{-3.5}$,][]{mathis1977size} from the minimum grain size of $0.1 \mathrm{~\mu m}$ to $1 \mathrm{~\mu m}$ grains. The minimum grain size corresponds to that of monomers, which are the building blocks of dust aggregates in protoplanetary discs \citep{dominik1997physics, tazaki2022size}. We set the fragmentation velocity to $v_\mathrm{frag} = 1 \mathrm{~m~s^{-1}}$ in line with recent experimental and observational works favouring relatively fragile grains \citep[e.g.,][]{gundlach2018tensile, musiolik2019contacts, jiang2024grain, ueda2024support}. We also test the case of more resistant dust grains with $v_\mathrm{frag} = 3 \mathrm{~m~s^{-1}}$ in Sect.~\ref{sec:results_v_frag}.

\subsection{Disentangling viscosity and local turbulence}
\label{sec:methods_viscosity_turbulence}

We separate the viscous disc $\alpha$-parameter into two components: (1) the global viscosity $\alpha_\mathrm{visc}$, governing the accretion of the disc material through angular momentum transport, and (2) a local turbulence parameter $\delta_\mathrm{turb}$ influencing the dust relative velocity and maximum size in the fragmentation-limited regime \citep{ormel2007closed, birnstiel2012simple} along with the radial and vertical dust diffusion \citep{semenov2011chemical, pinilla2021growing}. \refereed{In \texttt{DustPy}, this is done by setting both the radial and vertical dust diffusion parameters, $\delta_\mathrm{rad}$ and $\delta_\mathrm{vert}$, equal to $\delta_\mathrm{turb}$, which controls the turbulent collision velocities.} This approach allows us to move beyond the standard assumption that the disc turbulence and accretion is controlled by a unique dimensionless parameter $\alpha$ \citep{shakura1973black} which has been challenged by recent studies favouring weak local turbulence \citep[$\delta_\mathrm{turb} \approx 10^{-4}$, e.g.,][]{jiang2024grain, villenave2025turbulence} while the disc accretion rate \citep[$10^{-7}-10^{-10} \mathrm{~M_\odot~yr^{-1}}$][]{hartmann2016accretion, manara2020shooter} tends to be consistent with higher viscosity \citep{king2007accretion, rafikov2017protoplanetary}. Separating global accretion from local turbulence is even more important considering MHD winds could be efficient at driving the disc angular momentum transport, in which case one does not necessitate strong local turbulence to explain observed mass accretion rates \citep[][]{tabone2022secular}.

\subsection{Planetary gap model}
\label{sec:methods_bump_model}

\begin{figure}
    \centering
    \includegraphics[width=\columnwidth]{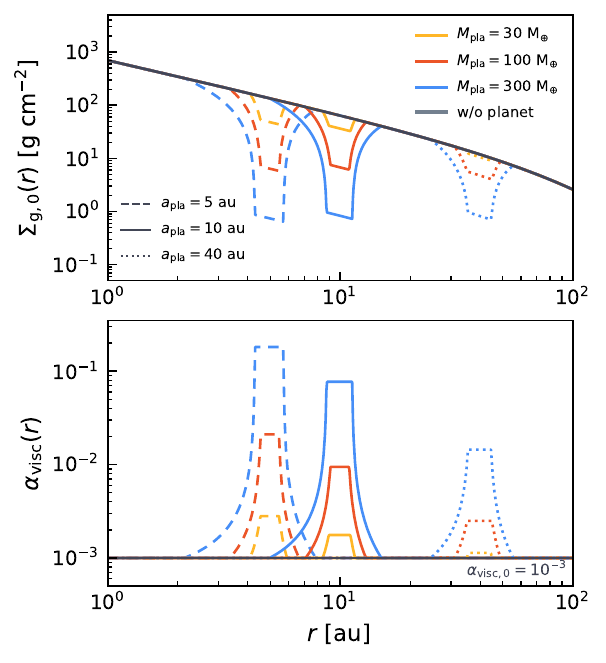}
    \caption{Radial profile of the gas surface density $\Sigma_\mathrm{g}$ (upper panels) and disc viscosity $\alpha_\mathrm{visc}$ (lower panels) including a dust trap caused by different planet mass and location, whose shape and depth is computed using empirical fits from \citet{kanagawa2017modelling}. We pick here our lowest value of the viscosity ($\alpha_\mathrm{visc, 0} = 10^{-3}$) which produces the deepest gaps in our parameter study. For comparison, $M_\mathrm{p} = 300 \mathrm{~M_\oplus}$ corresponds to an approximately Jupiter-mass planet.}
    \label{fig:SigmaG_AlphaVisc_gap}
\end{figure}

Concerning the dust trap, we assume it has a planetary origin\footnote{This assumption is only used to acquire a gap profile and a local pressure maximum. Our results remain valid for local pressure bumps originating from other processes.} with the planetary gap profile $F_\mathrm{gap}(r)$ (shape and depth) determined by empirical fits derived from 2D hydrodynamic simulations \citep[see equation 6 in][]{kanagawa2017modelling}. The planetary gap is thus characterised by the planet mass $M_\mathrm{p}$ and location $a_\mathrm{p}$, both assumed constant throughout the simulation. We impose the gap profile on the initial gas surface density profile $\Sigma_\mathrm{g,0}(r)$ and maintain it through the simulation by dividing the initial viscosity $\alpha_\mathrm{visc,0}$ as $\alpha_\mathrm{visc}(r) =\alpha_\mathrm{visc,0}/F_\mathrm{gap}(r)$, since the product of the gas surface density by the viscosity is constant in quasi steady-state \citep[a consequence of constant mass flux $\dot{M}$ across the gap, see][]{kanagawa2017modelling, duffell2020empirically, wafflard2023planet}. The local turbulence $\delta_\mathrm{turb}$ controlling particle size and diffusion is not affected by the gap profile, hence it is kept constant in our models. We show in Fig.~\ref{fig:SigmaG_AlphaVisc_gap} the radial profile of the initial gas surface density $\Sigma_\mathrm{g,0}(r)$ and viscosity parameter $\alpha_\mathrm{visc}(r)$ for different planet masses and locations, for a disc model where the unperturbed viscosity is $\alpha_\mathrm{visc,0} = 10^{-3}$. We see that planets, at an identical mass, tend to carve a deeper gap when they are closer to the star. \refereed{We also tested the empirical gap profile from \citet{duffell2020empirically}. Over the range of planet mass we explore (Sect.~\ref{sec:methods_param_space}), the main difference with \citet{kanagawa2017modelling} lies in slight changes to the shape of the gap edge. We nevertheless found no significant differences in the measured dust leakiness relative to \citet{kanagawa2017modelling} when using the gap profile of \citet{duffell2020empirically}.}

A limitation of using a one-dimensional code like \texttt{DustPy} is that we cannot self-consistently follow the planet position and model the detailed gas flow taking place in its surrounding, for which 2D and 3D hydrodynamic simulations are necessary \citep{weber2018characterising, haugbolle2019probing, petrovic2024material, vanclepper20253D, huang2025leaky, nicholson2025enhanced}. Such simulations find that small leaky grains pass the planetary gap on horseshoe orbits, sometimes entering the close vicinity of the planet \citep{petrovic2024material}, which may result in pebble accretion beyond the pebble isolation mass \citep[see figure 7 in][]{vanclepper20253D}. We discuss further the advantages and limitations of our approach compared to 2D/3D hydrodynamic simulations in Sect.~\ref{sec:discussion_limitation}.

Apart from the gas profile, we adopt a similar setup to \citet{stammler023leaky}, in which simulations are initialized by removing any dust grains located inside the planetary gap. \refereed{Specifically, we set up the initial dust surface density profile by removing dust located inside $r \leq 2a_\mathrm{p}$}. Removing the inner disc dust has several advantages. Firstly, it reduces the computational expense of simulations by several factors, hence allowing the exploration of a larger parameter space. Secondly, it simplifies the interpretation of our results, as any dust grains found inside the planetary gap must originate from the leaking of the dust trap. We demonstrate in Appendix~\ref{sec:appendix_removingdust} that this approach does not affect the analysis of leakiness from our simulations.

\subsection{Planetesimal formation}
\label{sec:methods_planetesimal_formation}

Dust traps may provide the necessary conditions for the onset of planetesimal formation through the streaming instability \citep[e.g.,][]{eriksson2020pebble}. Planetesimal formation would locally convert a fraction of the dust mass into planetesimals, hence decreasing the mass flux through the planetary gap. To properly model the leakiness of dust traps, it is thus necessary to include planetesimal formation in our model. Aligned with results from streaming instability simulations \citep{carrera2015form, li2021thresholds, lim2024streaming}, we trigger the formation of planetesimals in the dust trap if the midplane dust-to-gas ratio $\epsilon_\mathrm{d2g} = \rho_\mathrm{d}/\rho_\mathrm{g} \geq 1$. In that case, we follow the prescription of \citet{drazkowska2016close} and \citet{schoonenberg2018lagrangian}, and convert a fraction of the dust mass into planetesimals as
\begin{equation}
\label{eq:rate_planetesimal_formation}
    \dfrac{\partial \Sigma_\mathrm{pla}}{\partial t} = \zeta \Omega_\mathrm{K} \sum_{i} \Sigma_{\mathrm{d},i} \mathrm{St}_{i},
\end{equation}
where each particle size $i$ contributes inversely to its settling timescale $t_{\mathrm{set},i} = (\mathrm{St}_{i}\Omega_\mathrm{K})^{-1}$. We set the planetesimal formation efficiency\footnote{Streaming instability simulations by \citet{simon2016mass} find that about half of pebbles are turned into planetesimals over a few orbital periods, corresponding to $\zeta \approx 10^{-3}-10^{-2}$ \citep{drazkowska2016close}.} per settling timescale to $\zeta = 10^{-3}$, similarly to previous works \citep[e.g.,][]{andama2024which, lau2024sequential, zhao2025planetesimal}.

\subsection{Parameter space exploration}
\label{sec:methods_param_space}

Our aim is to provide a global overview of the impact of dust traps and their leakiness on the composition of the inner disc. Many parameters may influence the leakiness of traps, notably those influencing the size distribution, transport and collisional evolution of dust grains.

We chose to focus on a set of 4 fundamental parameters (see Sect.~\ref{sec:methods_viscosity_turbulence} and \ref{sec:methods_bump_model}): (1) the disc viscosity $\alpha_\mathrm{visc}$ that determines the inward motion of gas and small grains along with the gap depth, (2) the local turbulence $\delta_\mathrm{turb}$ that controls the dust particle size and the radial and vertical diffusion, (3) the planet mass $M_\mathrm{p}$, and (4) the planet location $a_\mathrm{p}$, both of which impact the gap profile (width and depth). To be consistent with the mass accretion rate observed in discs \citep[e.g.,][]{hartmann2016accretion, rafikov2017protoplanetary, manara2020shooter}, we pick 5 values for $\alpha_\mathrm{visc} \in [10^{-3}, 10^{-2}]$. 

The local gas turbulence in the protoplanetary disc is more complex to estimate. Considering that laboratory experiments favour fragile grains (see Sect.~\ref{sec:methods_dust_disc_model}), weak local turbulence $\delta_\mathrm{turb} \approx 10^{-4}$ is required to explain the dust particle size measured in discs with ALMA \citep{jiang2024grain, tong2026turbulence, luo2026almasurveygasevolution}. The analysis of the dust settling in a sample of 33 protoplanetary discs by \citet{villenave2025turbulence}, along with the gas emission turbulent broadening \citep{flaherty2017turbulence, teague2018turbulence}, also argue in favour of weak turbulence. In the meanwhile, the width of ALMA rings in the DSHARP survey \citep{dullemond2018dsharp, rosotti2020efficiency} provides hints of efficient radial and vertical diffusion, with a turbulence level as high as $\delta_\mathrm{turb} \gtrsim 10^{-3}$. However, if these rings have a planetary origin, the dust trap may host stronger turbulence due to the planet stirring material close to the gap edge \citep{binkert2023three}. In this case, turbulence levels inferred from ALMA rings may not be representative of the global value of $\delta_\mathrm{turb}$ in the protoplanetary disc. Overall, we pick 5 values across a larger range $\delta_\mathrm{turb} \in [10^{-4}, 10^{-2}]$. For each pair $(\alpha_\mathrm{visc}, \delta_\mathrm{turb})$, we test different planet masses $M_\mathrm{p} = [30, 100, 300]~\mathrm{M_\oplus}$ and locations $a_\mathrm{p} = [5, 10, 40]~\mathrm{au}$ (along with simulations without any planet). Taken together, our parameter scan represents over $300$ simulations.

\section{Results} 
\label{sec:results}

\subsection{The impact of dust traps} 
\label{sec:results_structured_vs_smooth}

\begin{figure*}
    \centering
    \includegraphics[width=\textwidth]{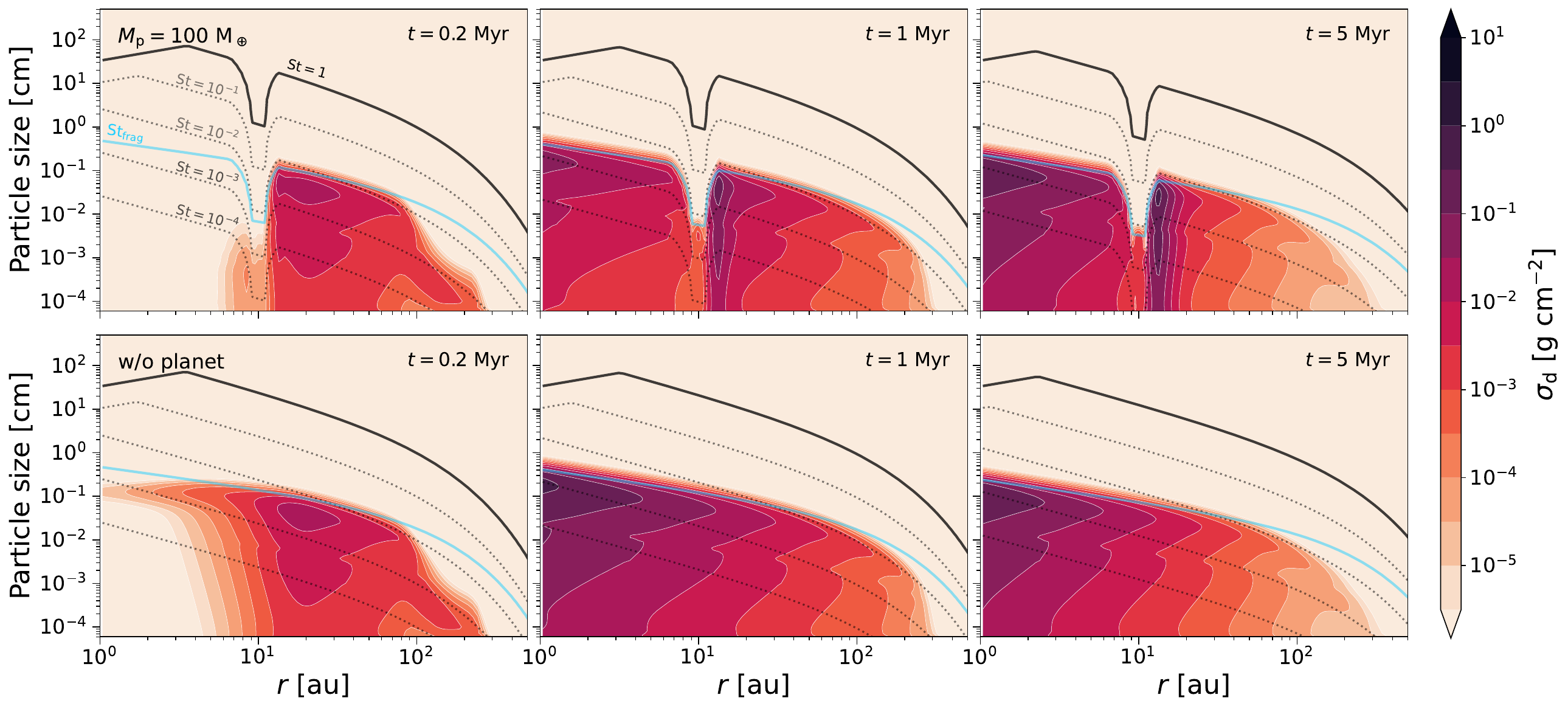}
    \caption{Vertically-integrated dust density distribution as a function of particle size for a disc with a gap carved by a $100 \mathrm{~M_\oplus}$ planet at $10 \mathrm{~au}$ (upper panels) compared to a smooth disc (lower panels). In the model showed here, the disc viscosity is set to $\alpha_\mathrm{visc} = 10^{-3}$ and the local turbulence to $\delta_\mathrm{turb} = 3.2 \times 10^{-4}$. The colour bar indicates the surface density of dust in a given size range at a given radius. The three panels represents different times of the simulation ($t = 0.2 \mathrm{~Myr}$, $t = 1 \mathrm{~Myr}$, and $t = 5\mathrm{~Myr}$). The black contours show different values of the Stokes number ($\mathrm{St} = [10^{-3}, 10^{-2}, 10^{-1}, 10^{0}]$) for visual guidance.}
    \label{fig:Fig2_SigmaD_gap_nogap}
\end{figure*}

\begin{figure}
    \centering
    \includegraphics[width=\columnwidth]{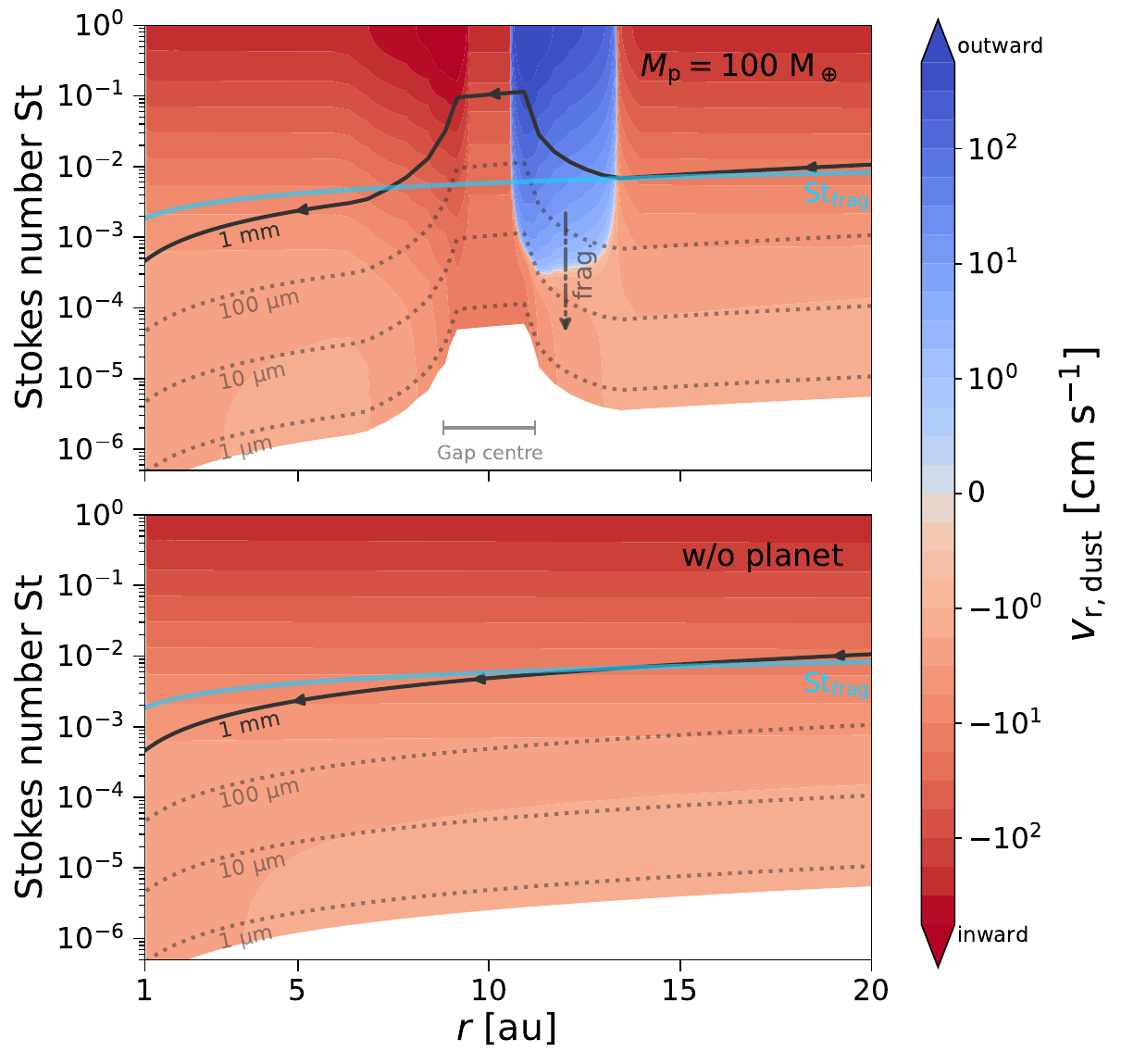}
    \caption{Dust velocity as a function of distance from the star and dust Stokes number in a disc hosting a $100 \mathrm{~M_\oplus}$ planet that has carved a gap at $10 \mathrm{~au}$ (upper panel) compared to a smooth disc (lower panel) for a global viscosity $\alpha_\mathrm{visc} = 10^{-3}$ and local turbulence $\delta_\mathrm{turb} = 3.2 \times 10^{-4}$. Negative velocity represents the inward motion of dust grains (red) while a positive velocity shows the outward motion (blue). The dark contours represent the variations in Stokes number for fixed particle size (from 1 $\mathrm{\mu m}$ to 1 mm). The Stokes number decreases monotonously in the smooth disc when approaching the star. In planet-hosting discs, low gas density in the planetary gap increases the Stokes number locally.}
    \label{fig:dust_velocity_vs_Stokes}
\end{figure}

In this section, we focus on a single, representative simulation to illustrate in greater details the impact of a local pressure maximum on dust evolution, and how a dust trap can be more or less leaky. We focus on the model of a disc with a $100 \mathrm{~M_\oplus}$ planet that has carved a gap at $10 \mathrm{~au}$, with a viscosity $\alpha_\mathrm{visc} = 10^{-3}$ and local turbulence $\delta_\mathrm{turb} = 3.2 \times 10^{-4}$. The fragmentation velocity is set to the nominal value $v_\mathrm{frag} = 1 \mathrm{~m~s^{-1}}$ (Sect.~\ref{sec:methods_dust_disc_model}). 

\subsubsection{Dust distribution and velocities}

We begin by presenting in Fig.~\ref{fig:Fig2_SigmaD_gap_nogap} the evolution of the dust density distribution at different times ($0.2\mathrm{~Myr}$, $1\mathrm{~Myr}$, $5 \mathrm{~Myr}$) for the planet-hosting disc (upper panels) as compared to a smooth disc (lower panels). As explained in Sect.~\ref{sec:methods_bump_model}, our initial conditions impose that no solid material is located inside the gap at the onset of the simulation. Fragile grains ($v_\mathrm{frag} = 1 \mathrm{~m~s^{-1}}$) coupled with weak local turbulence ($\delta_\mathrm{turb} = 3.2 \times 10^{-4}$) results in pebbles growing up to $\approx 1\, \mathrm{mm}$ around the planet location before being halted by fragmentation (blue line). This corresponds, in the dust trap outside of the planetary gap, to a Stokes number of $\mathrm{St_{frag}} \approx 10^{-2}$. By $t \lesssim 0.2 \mathrm{~Myr}$, outer disc solids have already passed through the gap on their way to the inner disc, though more slowly than in a smooth disc. The lower-left panel offers a visual on how the leakiness of dust traps works. Though mm-sized pebbles are blocked in the local pressure maximum, small grains are well coupled to the gas and move through the gap, before re-coagulating on the other side. Even though small grains make up only a minor fraction (here $\approx 20\%$) of the solid mass budget in the fragmentation-limited equilibrium, their constant replenishment by the fragmentation of pebbles gradually allow the mass stored in pebbles to leak through. After 1 Myr, the inner disc is filled with outer disc material. The dust distribution appears similar to that of a smooth disc. After 5 Myr, a substantial accumulation of dust is still present at the gap edge, which may promote planetesimal formation.

We then present in Fig.~\ref{fig:dust_velocity_vs_Stokes} the dust velocity\footnote{Note that dust diffusion, which is included in \texttt{DustPy}, also affects dust transport and will increase the leakiness of traps.} as a function of Stokes number for the planet-hosting disc (upper panel) as compared to a smooth disc (lower panel). The dust velocity is computed following
\begin{equation}
    \label{eq:dust_velocity}
    v_\mathrm{r, dust} = \Bigg[ v_\mathrm{gas} - 2 \eta v_\mathrm{K} \mathrm{St} \Bigg] \dfrac{1}{1 + \mathrm{St}^{2}}
\end{equation}
such that it depends on the gas velocity $v_\mathrm{gas}$ and a dust drift component proportional to the Stokes number and the midplane pressure gradient $\eta \propto -\,\partial \ln{P}/\partial \ln{r}$. For a smooth disc, dust grains move inward for all Stokes numbers (lower panel in Fig.~\ref{fig:dust_velocity_vs_Stokes}). The inward velocity is dominated by the gas speed for small grains and by the drift component for larger dust particles. In a planet-hosting disc, a local pressure maximum is present right outside the gap. As a result, large particles no longer drift inwards but rather move outwards towards the pressure maximum (blue region). For a $100 \mathrm{~M_\oplus}$ planet that has carved a gap at $10 \mathrm{~au}$, we find the critical Stokes number separating what is blocked from what leaks to be $\mathrm{St_{crit}} \approx 5 \times 10^{-4}$, corresponding at the gap edge to grains with a radius of $10 \mathrm{~\mu m}$ (see dashed contour). As fragmentation halts the growth of pebbles around $\mathrm{St_{frag}} \approx 10^{-2}$, most pebbles accumulate in the local pressure maximum in the blue region. However, as seen in Fig.~\ref{fig:Fig2_SigmaD_gap_nogap}, fragmentation gradually transfers mass from blocked pebbles (blue region, $\mathrm{St} > \mathrm{St_{crit}}$) to small grains (red region, $\mathrm{St} < \mathrm{St_{crit}}$), allowing a significant reservoir to leak. 

When approaching the gap, the gas density decreases such that grains with a constant size have higher Stokes number (see dark contours). For a $100 \mathrm{~M_\oplus}$ planet at $10 \mathrm{~au}$, the Stokes number increases in the planetary gap by an order of magnitude. As noted in \citet{weber2018characterising}, this restrains the size range of leaky grains as the increasing Stokes number close to the gap gives a higher outward velocity component to dust particles. Without that effect, particles approaching $100 \mathrm{~\mu m}$ could leak as well. This highlights that assuming a constant Stokes number to model the leakiness of dust traps can be inconsistent with the dust aerodynamics. Nevertheless, considering the mass exchange between pebbles and small grains with our full grain size distribution model, we see that this increase in Stokes number is not sufficient to prevent material from leaking (as is evident from Fig.~\ref{fig:Fig2_SigmaD_gap_nogap}).

Once in the gap, the radial velocity is directed inward across all Stokes number, hence they can no longer be halted on their way to the inner disc\footnote{As discussed in Sect.~\ref{sec:methods_bump_model}, our one-dimensional approach does not allow us to model the gas flow close to the planet, where small grains are most likely to cross \citep{petrovic2024material, vanclepper20253D}. This could change the velocity of grains once in the gap, though they should still move inwards.}. It is also interesting to see that in the gap, $\mathrm{St_{frag}} > \mathrm{St_{crit}}$. This means that even though the smallest grains can leak, they are able to re-coagulate inside the gap, and settle in a new fragmentation-limited equilibrium\footnote{The timescale to settle in a new fragmentation-limited equilibrium may be longer than the gap crossing time, especially for strong pressure maxima leading to low dust densities and collision rates in the gap. In that case, the particle size may be lower than the fragmentation barrier.} (also seen in Fig.~\ref{fig:Fig2_SigmaD_gap_nogap}), which corresponds to the same Stokes number $\mathrm{St_{frag}}$ as in the trap, though the actual particle size is lower. Overall, we can see that the gap is a highly dynamic environment for dust particles. Leaky grains do not simply move inward at a constant size, but fragment and re-coagulate continuously throughout their journey.

\subsubsection{Dust transport to the inner disc}

\begin{figure}
    \centering
    \includegraphics[width=\columnwidth]{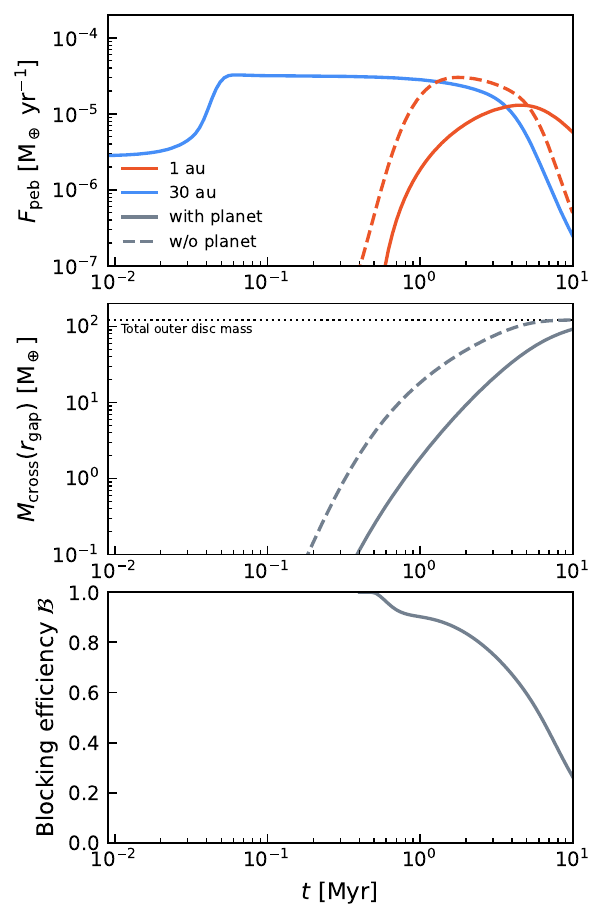}
    \caption{Key diagnostic quantities to quantify the leakiness of dust traps, here for the fiducial model of a disc with a $100 \mathrm{~M_\oplus}$ planet that has carved a gap at $10 \mathrm{~au}$ for a global viscosity $\alpha_\mathrm{visc} = 10^{-3}$ and local turbulence $\delta_\mathrm{turb} = 3.2 \times 10^{-4}$. \textit{Upper panel}: Pebble flux $F_\mathrm{peb}$ outside the planetary gap ($30 \mathrm{~au}$, blue) and reaching the inner disc at ($1 \mathrm{~au}$, red), for a planet-hosting disc (solid line) or smooth disc (dashed line). \textit{Middle panel}: Total solid mass $M_\mathrm{cross}$ that has crossed the radial location $r_\mathrm{gap} = 10 \mathrm{~au}$ since $t_0$ as a function of time for a planet-hosting disc (solid line) and smooth disc (dashed line). The dotted horizontal line indicates the total solid mass initially located outside of $10 \mathrm{~au}$. \textit{Lower panel}: Trap blocking efficiency $\mathcal{B}$ of the dust trap since $t_0$ as a function of time (see Eq.~\ref{eq:blockiness}).}
    \label{fig:F_peb_blockiness_onesim}
\end{figure}

To better quantify the impact of a gap on the material received by the inner disc, we show a few diagnostic quantities in Fig.~\ref{fig:F_peb_blockiness_onesim}, notably the pebble flux $F_\mathrm{peb}$ (upper panel) and the total mass of dust that has crossed the planetary gap, $M_\mathrm{cross}(r_\mathrm{gap})$ (middle panel). The latter is equivalent to the cumulative pebble flux since $t_0$ through the planetary gap location $r_\mathrm{gap}$ as
\begin{equation}
\label{eq:M_crossed}
    M_\mathrm{cross}(t, r_\mathrm{gap}) = \int_{t_0}^{t} F_\mathrm{peb}(t', r_\mathrm{gap})~\mathrm{dt'}.
\end{equation}
On the upper panel in Fig.~\ref{fig:F_peb_blockiness_onesim}, we see that the pebble flux is identical outside the planetary gap, but quite different in the inner disc. Initially, the pebble flux reaching the inner part of the smooth disc is greater, peaking at $F_\mathrm{peb} = 3 \times 10^{-5} \mathrm{~M_\oplus~yr^{-1}}$ before decaying around 5 Myr as the pebble reservoir empties out. In the case of a planet-hosting disc, the pebble flux is smaller, though it still reaches values $F_\mathrm{peb} \approx 10^{-5} \mathrm{~M_\oplus~yr^{-1}}$. This causes the pebble flux to be longer-lived, such that after 5 Myr the inner disc of the planet-hosting disc receives more solid material than the smooth disc \citep[see also][]{stammler023leaky}. In the middle panel in Fig.~\ref{fig:F_peb_blockiness_onesim}, we see that the total mass that has crossed the planetary gap (i.e., the cumulative flux over $r_\mathrm{gap} = 10 \mathrm{~au}$) is lower in a planet-hosting disc, but gets closer to the smooth disc with time. By 5 Myr, $75\%$ of the total solid mass initially located outside the planetary gap has crossed towards the inner disc.

We introduce now a new quantity we will use in this paper to make meaningful comparisons between the trapping efficiency of different dust traps: the trap blocking efficiency $\mathcal{B}$. The trap blocking efficiency quantifies the ability of a given dust trap to block the pebble flux over time, and is defined as
\begin{equation}
\label{eq:blockiness}
    \mathcal{B}(t) = 1 - \dfrac{M^\mathrm{gap}_\mathrm{cross}(t, r_\mathrm{gap})}{M^\mathrm{no~gap}_\mathrm{cross}(t, r_\mathrm{gap})}% - M_\mathrm{pla}^\mathrm{gap}(t),
\end{equation}
where $M^\mathrm{gap}_\mathrm{cross}$ and $M^\mathrm{no~gap}_\mathrm{cross}$ represent the total solid mass that has crossed $r = r_\mathrm{gap}$ since $t_0$ for a simulation with or without adding the planet-induced pressure perturbation in the disc profile (see middle panel in Fig.~\ref{fig:F_peb_blockiness_onesim}). With that definition, a blocking efficiency $\mathcal{B} = 0.75$ means that the dust trap has blocked $75\%$ of the solid material since the beginning of the disc lifetime compared to what drifted inward in a smooth disc. In the opposite, a blocking efficiency $\mathcal{B} = 0$ at a time $t$ means that by that time, a similar amount of dust made it to the inner disc compared to a smooth disc. By comparing each simulation to an identical control simulation performed without a planetary gap, we ensure that the blocking efficiency parameter $\mathcal{B}$ only quantifies the change in pebble flux that arises from the presence and leakiness of the gap. In contrast, varying parameters such as $\delta_\mathrm{turb}$ also alters the global dust dynamics in the disc, for example by modifying the particle size and inward drift motion, and hence the pebble flux even in a smooth disc. The comparison with the gap-free control simulation removes these global effects and ensures that the blocking efficiency parameter $\mathcal{B}$ only quantifies the contribution of the gap\footnote{For example, having a very low particle size would significantly reduce the radial drift of dust grains, hence lead to a small $M_\mathrm{cross}$ even though gaps are more leaky to small pebbles.}. With this definition of $\mathcal{B}$, we can also define a complementary parameter, namely the leaking efficiency $\mathcal{L} = 1 -\mathcal{B}$.

The lower panel in Fig.~\ref{fig:F_peb_blockiness_onesim} shows the evolution of the gap blocking efficiency vs. time. We can see that it quickly decreases with time, indicating that even if slowed, the planet-hosting disc is catching up on the amount of solid mass delivered to the inner disc. By the end of the simulation, the blocking efficiency is around $\mathcal{B} \approx 0.25$ indicating that $25 \%$ of the total drifting dust material ended up blocked by the planetary gap. Through this paper, we will compare dust traps focusing on their blocking efficiency after 5 Myr, as it represents typically disc lifetimes \citep[e.g.,][]{haisch2001disk}.

In the end, we can see that even though large pebbles, which dominate the solid mass reservoir in the disc, are blocked in the dust trap due to outward velocities (see Fig.~\ref{fig:dust_velocity_vs_Stokes}), the gradual transfer of mass from pebbles to grains by fragmentation allows material to leak through the dust trap (Fig.~\ref{fig:Fig2_SigmaD_gap_nogap}). In the disc model explored in this Section, this leaking is sufficient to allow $75 \%$ of the outer disc mass to make it through the planetary gap during the disc lifetime (Fig.~\ref{fig:F_peb_blockiness_onesim}).

\subsection{Parameter space exploration} 
\label{sec:results_paramspace}

\begin{figure*}
    \centering
    \includegraphics[width=\textwidth]{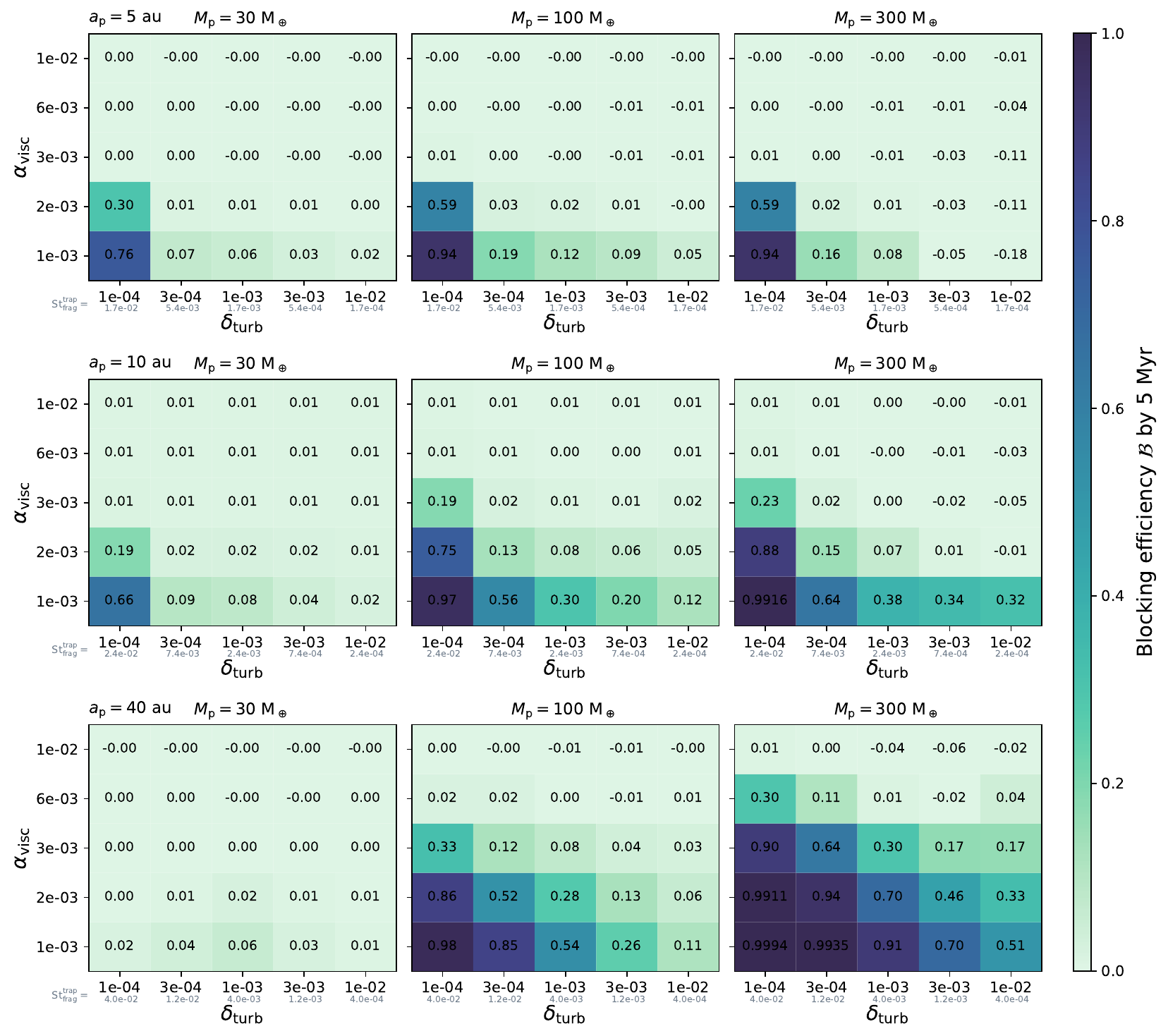}
    \caption{Trap blocking efficiency $\mathcal{B}$ (Eq.~\ref{eq:blockiness}) of a dust trap by $t=5\mathrm{~Myr}$ for a range of planet mass $M_\mathrm{p}$ ($30$, $100$, $300 \mathrm{~M_\oplus}$) and location $a_\mathrm{p}$  ($5$, $10$, $40 \mathrm{~au}$). The fragmentation velocity is set to $v_\mathrm{frag} = 1 \mathrm{~m~s^{-1}}$ (Sect.~\ref{sec:methods_dust_disc_model}). For each case, we run simulations for 25 pairs of viscosity $\alpha_\mathrm{visc}$ and local turbulence $\delta_\mathrm{turb}$. In total, $300$ simulations were used to build this figure. As discussed in Sect.~\ref{sec:results_structured_vs_smooth}, a blocking efficiency close to $0$ indicates a highly permeable dust trap, while a blocking efficiency approaching unity corresponds to a highly blocking trap. We indicate an estimate of the Stokes number of fragmentation-limited pebbles ($\mathrm{St_{frag}} = v_\mathrm{frag}^{2}/3\delta_\mathrm{turb}c_\mathrm{s}^2$) in the dust trap under the x-axis, as it varies with $\delta_\mathrm{turb}$ and position in the disc.}
    \label{fig:FigX_Blockiness_ap_Mp}
\end{figure*}

We now present our full parameter space exploration in Fig.~\ref{fig:FigX_Blockiness_ap_Mp}, representing the blocking efficiency $\mathcal{B}$ of the dust trap by $t = 5 \mathrm{~Myr}$ for each simulation. As discussed in Sect.~\ref{sec:methods_param_space}, we perform for each planet mass and location a total of $25$ simulations, representing pairs of disc viscosity $\alpha_\mathrm{visc}$ and local turbulence $\delta_\mathrm{turb}$. As the blocking efficiency is obtained by comparing with an identical simulation without a gap, each panel is built from $50$ \texttt{DustPy} simulations. As mentioned in Sect.~\ref{sec:methods_dust_disc_model}, the fragmentation velocity of grains is set to $v_\mathrm{frag} = 1 \mathrm{~m~s^{-1}}$, but we also explore more resistant grains ($v_\mathrm{frag} = 3 \mathrm{~m~s^{-1}}$) in Sect.~\ref{sec:results_v_frag}.

\subsubsection{Trap blocking efficiency vs. planet mass} 
\label{sec:results_Mp_variation}

We begin by analysing the impact of the planet mass (hence gap depth) on the blocking efficiency of dust traps. First, we focus on the central panel in Fig.~\ref{fig:FigX_Blockiness_ap_Mp} showing the blocking efficiency of a trap carved by a $100 ~\mathrm{M_\oplus}$ planet at $10 \mathrm{~au}$ for pairs of viscosity $\alpha_\mathrm{visc}$ and local turbulence $\delta_\mathrm{turb}$. We see that the majority of simulations have a blocking efficiency $\mathcal{B} \approx 0$, demonstrating that over a $5 \mathrm{~Myr}$ period, the inner disc has received a similar amount of solid material from the outer disc as compared to a smooth disc. We refer to such dust traps as \textit{fully permeable}. Typically, high viscosity $\alpha_\mathrm{visc} \geq 5 \times 10^{-3}$ results in fully permeable dust traps, and highly permeable from $\alpha_\mathrm{visc} \geq 3 \times 10^{-3}$. This is because the depth of planetary gaps decreases for higher viscosity \citep{kanagawa2017modelling}, increasing the size range of particles able to cross. Moreover, higher viscosity results in a stronger inward advection of the gas, making it more challenging for dust particles to be trapped (i.e., to have an outward velocity towards the pressure maximum, see Eq.~\ref{eq:dust_velocity} and Fig.~\ref{fig:dust_velocity_vs_Stokes}). As the viscosity decreases, the gap depth increases, which restrains the size of dust grains able to leak, and increases $\mathcal{B}$. Still, it is not sufficient to fully stop small grains from moving inwards with the gas. As we have seen in Sect.~\ref{sec:results_structured_vs_smooth}, fragmentation gradually transfers the mass from blocked pebbles to leaky grains, resulting in all cases to $\mathcal{B} < 1$.

Similarly to the viscosity, stronger local turbulence makes leakier dust traps. This is due to the combination of two effects. First, strong turbulence limits the maximum size of pebbles, as in the fragmentation-limited regime we have $a_\mathrm{max} \propto \delta_\mathrm{turb}^{-1}$. Therefore, stronger local turbulence results in pebbles with smaller size and Stokes number (see x-axis in Fig.~\ref{fig:FigX_Blockiness_ap_Mp}), which are more coupled to the inward gas flow. Second, local turbulence controls the diffusion of dust particles, which fights against the pile-up of dust in the trap. Even at low viscosity, strong local turbulence (lower-right corner of the panel) leads to strong leakiness of the dust trap. In the opposite limit, the most efficient blocking occurs for discs with low viscosity and weak local turbulence (lower-left corner of the panel) in which case $\mathcal{B} =0.97$ by 5 Myr. In that case, more than an order of magnitude of outer disc solids remain trapped, as compared to a smooth disc.

We can now discuss how the blocking efficiency evolves for varying planet mass, comparing panels in different columns in Fig.~\ref{fig:FigX_Blockiness_ap_Mp}. For a lower planet mass ($30 \mathrm{~M_\oplus}$, left panels), the blocking efficiency decreases for all values of viscosity and local turbulence. Effectively, only the lowest pair of viscosity and turbulence slows down dust transport, though it is limited to a few factors as compared to a smooth disc. For a more massive planet ($300 \mathrm{~M_\oplus}$, right panels) comparable to Jupiter, the blocking efficiency increases and a larger fraction of dust traps in our parameter space begin to slow down the transport of pebbles. We can also see in the upper-right corner the appearance of a particular regime where $\mathcal{B} < 0$, which we call \textit{super-leaky}. In that case, not only is the gap fully permeable, but it also gives an inward kick to all particles. This is because in the gap centre, the inward gas velocity is higher (see Fig.~\ref{fig:dust_velocity_vs_Stokes}), which transports particles faster across the gap. Acting as an highway for the dust population, it gives a stronger pebble flux in the planet-hosting disc as compared to a smooth disc.

\subsubsection{Trap blocking efficiency vs. gap location} 
\label{sec:results_ap_variation}

We now focus on the impact of the planet location on the dust trap blocking efficiency , comparing panels in different rows in Fig.~\ref{fig:FigX_Blockiness_ap_Mp}.

When varying the planet location, two effects compete in shaping the blocking efficiency of the dust trap. On the one hand, for an identical planet mass, the gap is deeper closer to the star, and weaker further out (see Fig.~\ref{fig:SigmaG_AlphaVisc_gap}). On the other hand, the properties of dust particles change with distance from the star. For a given particle size, the Stokes number is lower in the inner disc because of the higher gas density, such that it is more challenging to trap grains there (Fig.~\ref{fig:dust_velocity_vs_Stokes}). These two effects result in opposite trends on how the blocking efficiency of the trap varies with the planet location.

For a massive planet ($M_\mathrm{p} \geq 100 \mathrm{~M_\oplus}$), the second process dominates. Even though the planet carves a weaker gap in the outer disc, dust grains have higher Stokes number and hence are easier to trap. In fact, the most blocking dust trap found in our parameter study ($\mathcal{B} = 0.9994$ by 5 Myr) is in the case of a Jupiter-mass planet at $40 \mathrm{~au}$, for the lowest viscosity and turbulence, in which case only the smallest grains close to monomer size are leaking \citep[see also][]{pinilla2024survival}. Meanwhile, the blocking efficiency of a dust trap at $5 \mathrm{~au}$ is much lower, with almost all dust traps in our parameter study being highly permeable. It is also interesting to note that at $5 \mathrm{~au}$, the $100 \mathrm{~M_\oplus}$ planet appears slightly more blocking than the $300 \mathrm{~M_\oplus}$ planet. As described in Sect.~\ref{sec:results_Mp_variation}, this is because more massive planets carve deeper gaps, which also gives a stronger inward velocity to leaky grains (super-leaky regime). In fact, in the upper-right panel ($300 \mathrm{~M_\oplus}$ at $5 \mathrm{~au}$), low $\alpha_\mathrm{visc} = 10^{-3}$ but strong $\delta_\mathrm{turb} = 10^{-2}$ result in a deep gap with small particle size, overall leading a super-leaky trap with $\mathcal{B} = -0.18$ (the most super-leaky trap in our parameter study).

For low-mass planets ($M_\mathrm{p} = 30 \mathrm{~M_\oplus}$), the first process dominates and the trend is opposite. The gap is so weak in the outer disc (Fig.~\ref{fig:SigmaG_AlphaVisc_gap}) that its presence barely influence dust transport. Overall, at $40 \mathrm{~au}$ (lower-left panel) the dust trap is fully permeable for all values of viscosity and local turbulence.

To summarize, inner dust traps are rather leaky, while outer disc traps can either be much leakier or blockier depending on the planet mass. In some cases, solid mass flows inward more efficiently in planet-hosting discs compared to smooth discs. For planets at $a_\mathrm{p} \leq 10 \mathrm{~au}$ and up to a Jupiter-mass, protoplanetary discs with $\alpha_\mathrm{visc} \geq 3 \times 10^{-3}$ and/or $\delta_\mathrm{turb} \geq 10^{-3}$ transport over $5 \mathrm{~Myr}$ a similar amount of mass to the inner disc as compared to a smooth disc. The most extreme blocking occurs for low viscosity, weak local turbulence, and $M_\mathrm{p} \geq 100 \mathrm{~M_\oplus}$, reaching $\mathcal{B} > 0.95$. As we will see in Sect.~\ref{sec:results_planetesimalformation}, this is also related to planetesimal forming efficiently at weak local turbulence. In the parameter space we explore, complete blocking ($\mathcal{B} = 1$) is never achieved. We will discuss what it takes to reach this regime in Sect.~\ref{sec:discussion_leaking_regime}.

\subsubsection{Trap blocking efficiency for resistant grains: $v_\mathrm{frag} = 3 \mathrm{~m~s^{-1}}$}
\label{sec:results_v_frag}

\begin{figure}
    \centering
    \includegraphics[width=\columnwidth]{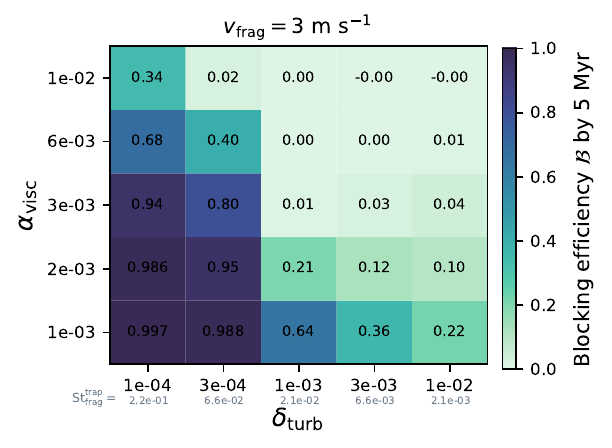}
    \caption{Trap blocking efficiency of a dust trap carved by a $100 \mathrm{~M_\oplus}$ planet at $10 \mathrm{~au}$ by $t=5\mathrm{~Myr}$ for resistant dust grains ($v_\mathrm{frag} = 3 \mathrm{~m~s^{-1}}$), for a range of viscosity $\alpha_\mathrm{visc}$ and local turbulence $\delta_\mathrm{turb}$. We indicate an estimate of the Stokes number of fragmentation-limited pebbles ($\mathrm{St_{frag}} = v_\mathrm{frag}^{2}/3\delta_\mathrm{turb}c_\mathrm{s}^2$) in the dust trap under the x-axis, similarly to Fig.~\ref{fig:FigX_Blockiness_ap_Mp}.}
    \label{fig:FigX_Blockiness_vf3}
\end{figure}

We now present in Fig.~\ref{fig:FigX_Blockiness_vf3} the blocking efficiency of dust traps by 5 Myr for the case of more resistant dust particles ($v_\mathrm{frag} = 3 \mathrm{~m~s^{-1}}$) for a trap caused by a $100 \mathrm{~M_\oplus}$ planet at $10 \mathrm{~au}$. In the fragmentation-limited equilibrium, this corresponds to an increase in pebble size and Stokes number of approximately an order of magnitude, as shown on the x-axis \citep[$a_\mathrm{max} \propto v_\mathrm{frag}^2$][]{birnstiel2011dust}.

Compared to fragile grains ($v_\mathrm{frag} = 1 \mathrm{~m~s^{-1}}$, middle panel in Fig.~\ref{fig:FigX_Blockiness_ap_Mp}) the blocking efficiency is much higher, especially for the lower turbulence range ($\delta_\mathrm{turb} < 10^{-3}$). As we will discuss in Sect.~\ref{sec:results_planetesimalformation}, this is because a larger fragmentation velocity results in more trapping and a more vertically-settled midplane, triggering planetesimal formation in a greater number of dust traps. Note, however, that larger fragmentation velocity coupled with weak turbulence results in very large pebbles ($\mathrm{St_{frag}} \approx 0.1$ in the trap, corresponding to a few centimetres in size), which is not in agreement with observations \citep{jiang2024grain, tong2026turbulence, luo2026almasurveygasevolution}.

\subsection{Forming planetesimals in dust traps} 
\label{sec:results_planetesimalformation}

\begin{figure*}
    \centering
    \includegraphics[width=\textwidth]{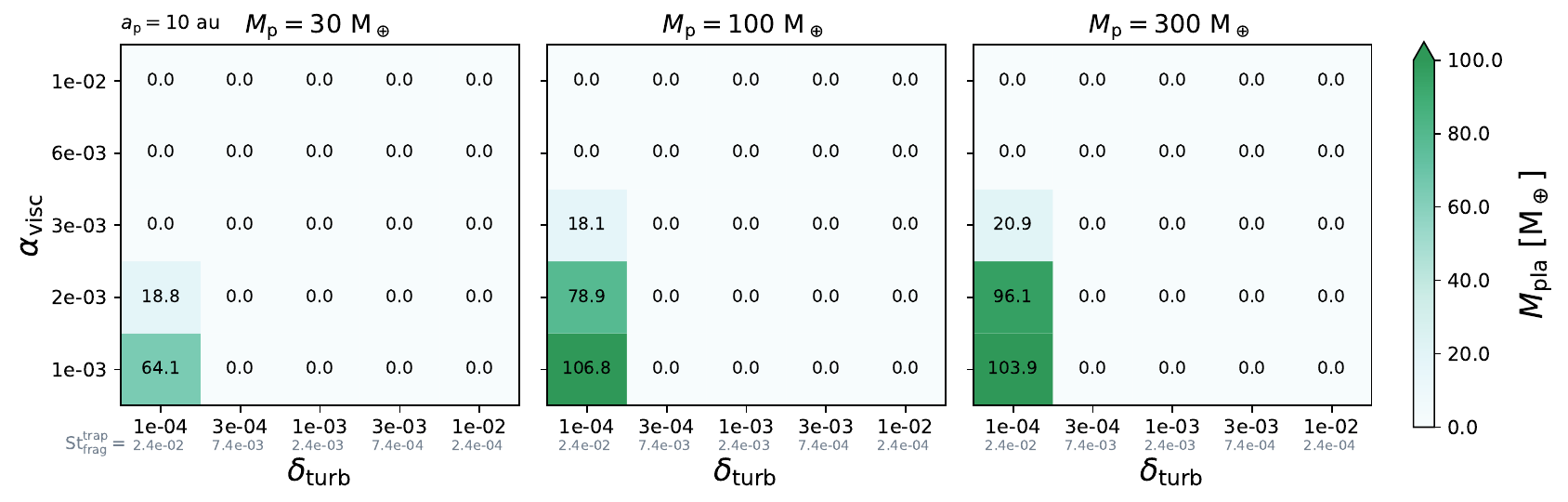}
    \caption{Mass of planetesimals formed in the dust trap generated by a planet at $10 \mathrm{~au}$ by the end of our simulations for a range of viscosity $\alpha_\mathrm{visc}$, local turbulence $\delta_\mathrm{turb}$ and planet mass $M_\mathrm{p}$. We see that under specific conditions, dust traps can be strong enough to form planetesimals. In that case, a large fraction of the available outer disc dust mass (see Fig.~\ref{fig:F_peb_blockiness_onesim}) may be transferred into planetesimals.}
    \label{fig:FigX_Mpla_end}
\end{figure*}

As mentioned in Sect.~\ref{sec:methods_planetesimal_formation}, we include the formation of planetesimals in our simulations, provided that the accumulation of dust in the trap is sufficient to raise the dust-to-gas ratio above unity. Thanks to our broad parameter study, we have access to a large sample of dust trap strength and shape, allowing us to determine which are strong enough to satisfy these conditions. We present in Fig.~\ref{fig:FigX_Mpla_end} the total mass of planetesimals formed by the end of the simulation (10 Myr) for different planet masses at $10 \mathrm{~au}$ over a range of viscosity $\alpha_\mathrm{visc}$ and local turbulence $\delta_\mathrm{turb}$.

We see that most traps in our sample do not form planetesimals. This is either due to the dust trap being too leaky, or vertical diffusion prohibiting the concentration of dust in the midplane, both of which limiting the dust-to-gas ratio below the critical threshold  $\epsilon_\mathrm{d2g, crit} \geq 1$ (see Sect.~\ref{sec:methods_planetesimal_formation}). For stronger dust traps and weak turbulence, planetesimals form with varying efficiency, from about $10$ to $100 \mathrm{~M_\oplus}$ of planetesimals. As also highlighted in previous works \citep[e.g.,][]{drazkowska2018buildup, miller2021formation, zhao2025planetesimal}, weak local turbulence ($\delta_\mathrm{turb} \sim 10^{-4}$) is highly beneficial to planetesimal formation. This is because it simultaneously increase the trapping of pebbles (increasing their size and lowering radial diffusion) while also reducing vertical mixing, boosting the dust density in the midplane. The disc viscosity $\alpha_\mathrm{visc}$ also impacts planetesimal formation as it affects the depth of the gap. Even for weak turbulence, strong viscosity $\alpha_\mathrm{visc} \geq 5 \times 10^{-3}$ prevents planetesimal formation.

\begin{figure}
    \centering
    \includegraphics[width=\columnwidth]{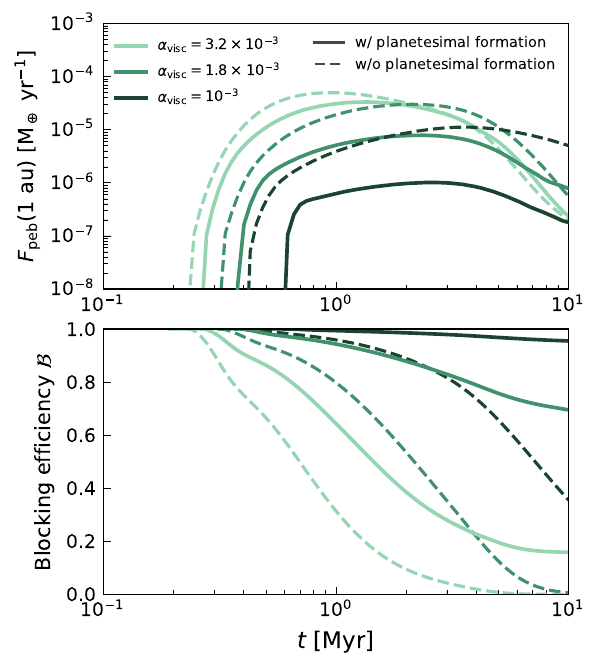}
    \caption{Pebble flux at 1 au and dust trap blocking efficiency $\mathcal{B}$ as a function of time for simulations with or without planetesimal formation. We focus on the case of a gap carved by a $100 \mathrm{~M_\oplus}$ planet at $10 \mathrm{~au}$, in which case only three simulations reach the requirements for planetesimal formation ($\delta_\mathrm{turb} = 10^{-4}$ and $\alpha_\mathrm{visc} \in [10^{-3}, 3.2 \times 10^{-3}]$, see Fig.~\ref{fig:FigX_Mpla_end}). We see that highly blocking dust traps ($\mathcal{B} > 0.9$) can only be achieved thanks to planetesimal formation. If planetesimals did not form, a dust trap in the same disc conditions would be much leakier.}
    \label{fig:Fig_Fleak_with_without_planetesimals}
\end{figure}

To quantify how much the onset of planetesimal formation reduces the leaking of dust grains through the dust trap, we performed additional simulations turning off planetesimal formation (i.e., setting the formation efficiency $\zeta = 0$, see Eq.~\ref{eq:rate_planetesimal_formation}). We present this comparison in Fig.~\ref{fig:Fig_Fleak_with_without_planetesimals}, which shows the pebble flux reaching $1 \mathrm{~au}$ (upper panel) and dust trap blocking efficiency (lower panel) as a function of time, similarly to Fig.~\ref{fig:F_peb_blockiness_onesim}. We focus on the 3 simulations that formed planetesimals for a planetary gap carved by a $100 \mathrm{~M_\oplus}$ planet at $10 \mathrm{~au}$. We see that the pebble flux is lowered when planetesimal formation is included, from a few factors up to two orders of magnitude depending on the amount of planetesimals formed. The blocking efficiency is also much larger when planetesimal formation is included. \refereed{We note that the pebble flux may be further reduced considering pebble accretion by planetesimals in the dust trap \citep[see][]{lau2022rapid, lau2024sequential, lau2025formation}, although only the largest planetesimals formed by the streaming instability are expected to undergo significant growth by pebble accretion \citep{lorek2026from}.}

This highlights one of the key findings of our work: highly blocking dust traps are so because they satisfy the conditions for planetesimal formation, in which case dust is converted into larger bodies. This is also one of the reason why higher fragmentation velocity (Fig.~\ref{fig:FigX_Blockiness_vf3}) results in much more highly blocking dust traps: the higher pebble size leads to a more vertically-settled midplane layer, facilitating planetesimal formation. If planetesimal formation did not happen in dust traps, for example because SI is inactive at pressure gradient near zero \citep{abod2019mass, eriksson2020pebble}, or because planet-induced turbulence stirs solids in the trap \citep{binkert2023three}, highly blocking dust traps with $\mathcal{B} > 0.9$ would be much more challenging to achieve\footnote{Moreover, solid mass stored in planetesimals may be returned to small dust grains via ablation, in which case the leaking flux would be higher \citep{eriksson2021fate}. This effect is not included in this work.}. Interestingly, the impact of planetesimal formation in reducing the leakiness of dust traps was not identified in previous studies based on simpler dust evolution models \citep{kalyaan2023effect}. As we will discuss further in Sect.~\ref{sec:discussion_JWST}, our parameter study then finds that it is planetesimal formation rather than the dust trap itself which can dramatically alter the inner disc composition (e.g., oxygen depletion).

\subsection{Comparisons with simpler dust evolution models}
\label{sec:results_2pop_vs_DustPy}

\begin{figure}
    \centering
    \includegraphics[width=\columnwidth]{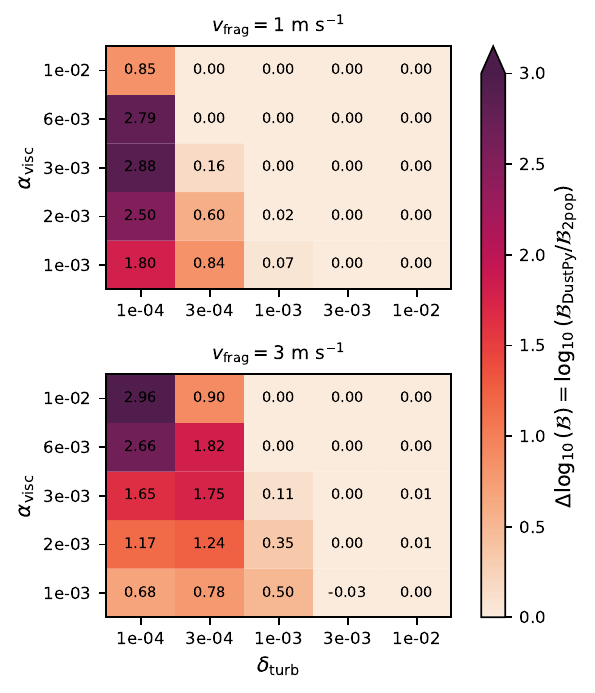}
    \caption{Comparison between the dust trap blocking efficiency found by our full grain size distribution model based on \texttt{DustPy}, $\mathcal{B}_\mathrm{DustPy}$, and the simpler two-population algorithm, $\mathcal{B}_\mathrm{2pop}$ \citep{birnstiel2012simple} here incorporated within chemcomp \citep{schneider2021how}, for identical disc models with fragile grains (upper panel) and resistant grains (lower panel). We see that simpler dust evolution models can overestimate by orders of magnitude the amount of mass blocked in dust traps, hence they may significantly underestimate the flux of outer disc material reaching the inner disc.}
    \label{fig:twopop_vs_DustPy}
\end{figure}

As \texttt{DustPy} does not directly evolve the abundance of volatile species, other models based on the two-population algorithm \citep[\texttt{two-pop-py}, see][]{birnstiel2012simple} have been mainly used in the literature when investigating the impact of dust transport on inner disc composition \citep{booth2017chemical, booth2019planet, mah2023close, lienert2024changing,houge2025burned, houge2025smuggling}, including in the presence of dust traps \citep{kalyaan2023effect, mah2024mind, sellek2025co2}. However, the two-population algorithm was built to model the evolution of smooth discs \citep{birnstiel2012simple}, such that it is not clear how well it performs in modelling the transport of dust in the presence of a dust trap. 

To investigate the performance of simpler dust evolution models, we compare in Fig.~\ref{fig:twopop_vs_DustPy} the blocking efficiency computed by \texttt{DustPy}, $\mathcal{B}_\mathrm{DustPy}$, with that of the two-population algorithm, $\mathcal{B}_\mathrm{2pop}$ \citep[incorporated within \texttt{chemcomp},][]{schneider2021how}, in the presence of a $100 \mathrm{~M_\oplus}$ at $10 \mathrm{~au}$, for fragile (upper panel) and resistant dust grains (lower panel). The difference $\Delta \mathcal{B}$ between $\mathcal{B}_\mathrm{DustPy}$ and $\mathcal{B}_\mathrm{2pop}$ is computed as 
\begin{equation}
    \Delta \log_{10}(\mathcal{B}) = \log_{10} \Bigg({\dfrac{\mathcal{B}_\mathrm{2pop}}{\mathcal{B}_\mathrm{DustPy}}}\Bigg),
\end{equation}
such that $\Delta \log_{10}(\mathcal{B}) = 0$ means that both models agree, while larger positive or negative values indicate orders of magnitude differences.

%\footnote{One model ($\alpha_\mathrm{visc} = 10^{-3}$, $\delta_\mathrm{turb} = 3.2 \times 10^{-3}$, $v_\mathrm{frag} = 3 \mathrm{~m~s^{-1}}$) has $\Delta \mathcal{B} \lesssim 0$. This is because \texttt{two-pop-py} estimates the trap to be more \textit{super-leaky} than \texttt{DustPy}.}

For strong turbulence ($\delta_\mathrm{turb} \gtrsim 10^{-3}$), the difference in blocking efficiency is close to zero, indicating that \texttt{two-pop-py} performs relatively well. This is because in these simulations, the dust trap is quite permeable (see Sect.~\ref{sec:discussion_leaking_regime}), resembling the case of a smooth disc which the two-population algorithm was designed for. For lower turbulence ($\delta_\mathrm{turb} \lesssim 10^{-3}$), however, we see that \texttt{two-pop-py} consistently underestimates the leaking of dust traps, from a few factors up to $3$ orders of magnitude. The difference is worst when the overestimation of the trapped solid mass by \texttt{two-pop-py} results in planetesimal formation, whereas the corresponding \texttt{DustPy} models remain below the threshold for the onset of the streaming instability. Overall, our parameter study with \texttt{DustPy} shows that dust traps are leakier than previously thought, on a broader parameter space.

These significant differences come from the fact that in \texttt{two-pop-py}, the dust population (small grains and pebbles) drifts at a common mass-weighted average speed $\overline{v_\mathrm{d}}$ \citep{birnstiel2012simple}, which is dictated by the velocity of the largest pebbles. Once pebbles are blocked, \texttt{two-pop-py} mimics the leakiness by allowing mass to be transferred across the gap by diffusion \citep[see also Eq.11 in][]{sellek2025co2}. This transition from drift to diffusion-dominated leaking can result in sharp variations in the pebble flux estimated by \texttt{two-pop-py} (see Appendix~\ref{sec:appendix_twopop_DustPy}). Moreover, for intermediate to low turbulence, the diffusive flux is small which leads to a significant overestimate of the amount of mass blocked in the dust trap. In the meanwhile, \texttt{DustPy} considers a full dust size distribution, including the mass transfer between different size bins through coagulation and fragmentation, and self-consistently computes the dust velocity as a function of particle sizes $v_\mathrm{d}(a)$. It can thus capture the leakiness and its variation across the parameter space more precisely.

Given that previous models linking the impact of outer disc dust traps on the inner disc composition were based on \texttt{two-pop-py} \citep[][]{kalyaan2023effect, mah2024mind, sellek2025co2}, it is possible that they underestimated the leaking flux, hence the delivery of materials (e.g., oxygen in the form of water ice) from the cold outer regions. We will come back to that in Sect.~\ref{sec:discussion_JWST}.

\subsection{Pebble flux and volatile delivery to the inner disc} 
\label{sec:results_M_H2O}

\begin{figure*}[h!]
    \centering
    \includegraphics[width=\textwidth]{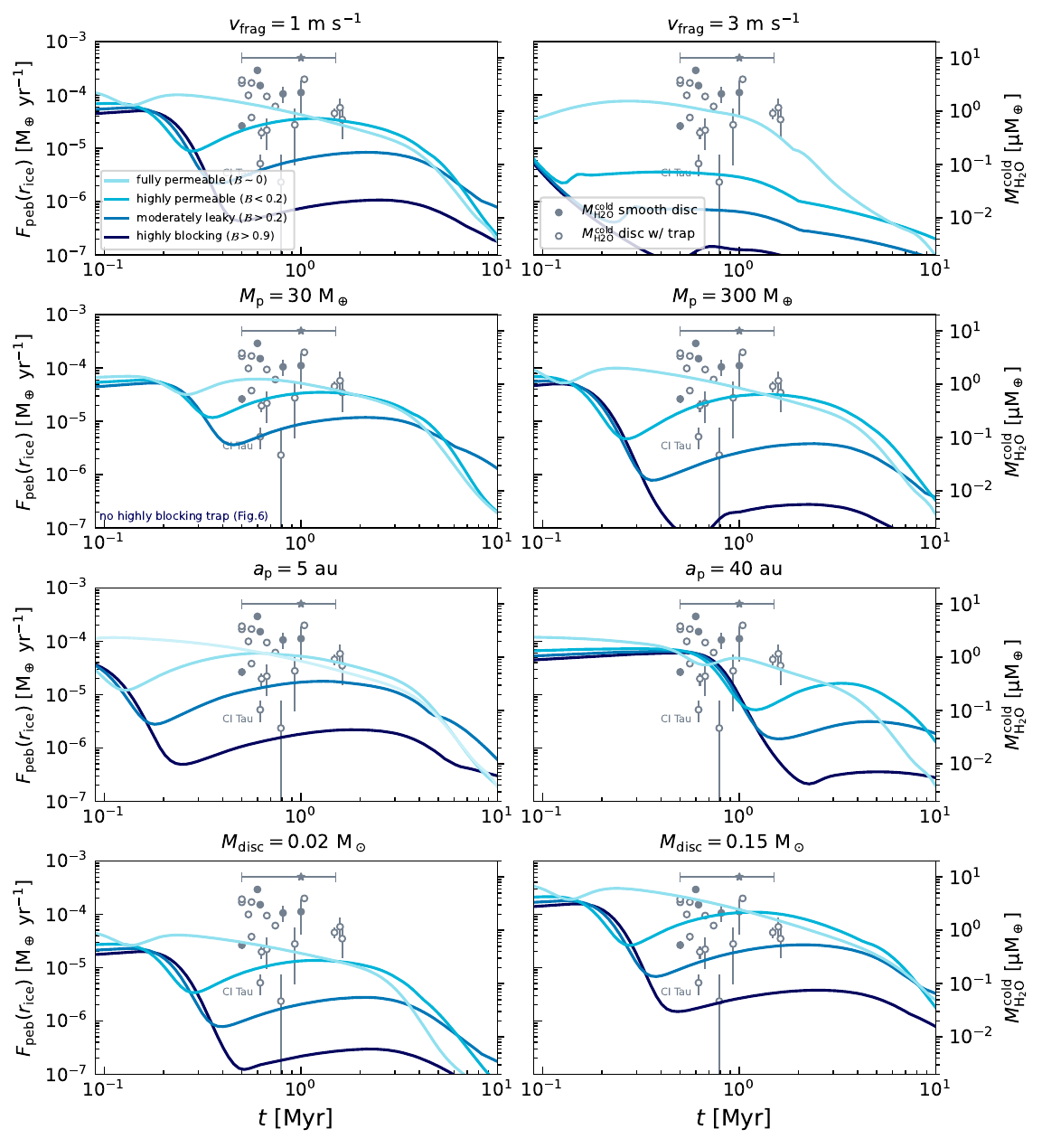}
    \caption{Pebble flux through the iceline (left axis) and observable cold water vapour mass (right axis) vs. time for different dust trap leakiness and parameters. The different colours indicate simulations with $\delta_\mathrm{turb} = 10^{-4}$ and varying viscosity from $\alpha_\mathrm{visc} = 10^{-3}$ (most blocking, dark blue) to $\alpha_\mathrm{visc} = 10^{-2}$ (fully permeable, light blue). We present these four cases because as seen in Fig.~\ref{fig:FigX_Blockiness_ap_Mp}, varying the viscosity across this column samples well the different leakiness regimes. Note that for weak traps (e.g., with $M_\mathrm{p} = 30 \mathrm{~M_\oplus}$), even the most blocking trap is quite leaky. Grey dots indicate measurements of the vapour reservoir made with JWST \citep[see Figure 2 in][]{krijt2025cosmic} and converted in pebble flux following Eq.~\ref{eq:F_peb_to_M_cold}. The title of each panel indicates the parameter that is varied from the standard fiducial values: $v_\mathrm{frag} = 1 \mathrm{~m~s^{-1}}$, $M_\mathrm{p} = 100 \mathrm{~M_\oplus}$, $a_\mathrm{p} = 10 \mathrm{~au}$ and $M_\mathrm{disc} = 0.05 \mathrm{~M_\odot}$. The grey star label and horizontal line indicate an estimate of the error on the disc age \citep[$\sim 0.5 \mathrm{~Myr}$,][]{krijt2025cosmic}.}
    \label{fig:FigY_Mvap_Fpeb}
\end{figure*}

We now present how different dust trap strengths affect the amount of material being delivered to the inner disc, focusing on water ice, and how it relates to observations made with JWST.

Infrared spectra gathered with JWST do not probe the full water reservoir, but a small observable fraction located in the disc upper atmosphere \citep{woitke2018modelling}. More precisely, spectral lines of colder water seem to directly probe the ongoing pebble flux through the water iceline \citep{banzatti2023JWST, banzatti2025water, vlasblom2025understanding}. Using the approach of \citet{romero2024retrieval}, it is possible to estimate analytically the cold water vapour reservoir $M_\mathrm{H_2O}^\mathrm{cold}$ from the pebble flux through the water iceline $F_\mathrm{peb}(r_\mathrm{ice})$, which is derived from our simulations. The water iceline is defined as the location where $T = 150 \mathrm{~K}$ and is located at $r_\mathrm{ice} = 1.1 \mathrm{~au}$. Following \citet{romero2024retrieval} and \citet{krijt2025cosmic}, the observed cold water vapour mass is computed as
\begin{equation}
\label{eq:F_peb_to_M_cold}
    M_\mathrm{H_2O}^\mathrm{cold}(t) = F_\mathrm{peb}(r_\mathrm{ice},t)t_\mathrm{H_2O} \dfrac{f_\mathrm{H_2O, ice}}{\eta},
\end{equation}
where $f_\mathrm{H_2O,ice} = 0.2$ is the mass fraction of water ice on pebbles \citep{patzold2016homogeneous}, $t_\mathrm{H_2O} = 100 \mathrm{~yr}$ is the timescale on which newly delivered water vapour is removed by chemistry \citep[e.g., by FUV photodissociation,][]{vlasblom2025understanding} or advection, and $\eta$ is a correction accounting for the fraction of water vapour hidden under the optically thick layer \citep{houge2025smuggling, sellek2025co2}. Similarly to previous works, we assume a constant value $\eta_\mathrm{obs} = 10^{3}$ \citep{romero2024retrieval, krijt2025cosmic} but we note that this parameter is both a function of time and varies with dust evolution \citep[see figure 5 in][]{houge2025smuggling}.

We present in Fig.~\ref{fig:FigY_Mvap_Fpeb} the pebble flux through the water iceline $F_\mathrm{peb}(r_\mathrm{ice})$ and cold water vapour mass as a function of time for different fragmentation velocities, planet masses, planet locations and initial disc masses. In each panel, we present simulations with varying degree of trap leakiness, corresponding to weak local turbulence $\delta_\mathrm{turb} = 10^{-4}$ with viscosity ranging from $\alpha_\mathrm{visc} = 10^{-3}$ (most blocking, dark blue) to $\alpha_\mathrm{visc} = 10^{-2}$ (fully permeable, light blue). As seen in Fig.~\ref{fig:FigX_Blockiness_ap_Mp}, this allows to sample well the different leakiness regimes. We also show the estimates of $M_\mathrm{H_2O}^\mathrm{cold}$ made from JWST observations \citep[see Figure 2 in][]{krijt2025cosmic}. As we focus here on a smaller subset of our parameter study, we ran the simulations for a full disc instead of removing the dust reservoir inside the planetary gap, in order to resolve the pebble flux during early stages (see Sect.~\ref{sec:methods} and Appendix~\ref{sec:appendix_removingdust}).

We see that in the early stages, the pebble flux reaching the water iceline is unaffected by the presence of the outer dust trap, as the dust reservoir initially inside the trap drifts inward like in a smooth disc. It takes some time \citep[$t_\mathrm{delay}$ as introduced in][]{krijt2025cosmic} for the pebble flux to decrease, by an amount sensitive to the blocking efficiency of the dust trap. Only after that time is the composition of the inner disc impacted by the presence of the outer dust trap. We notice that this delay decreases for inner traps, as in this case most of the dust reservoir is initially located outside the gap. Similarly, higher fragmentation velocities result in larger pebbles drifting faster to the inner disc, hence decreasing the delay.

Fully permeable traps (light blue lines) typically reach $F_\mathrm{peb}(r_\mathrm{ice}) \approx 10^{-4} \mathrm{~M_\oplus~yr^{-1}}$, which is similar to previous works on smooth discs \citep[e.g.,][]{drkazkowska2021dust, williams2025fuelled}. The pebble flux then decreases by a few factors when increasing the trap blocking efficiency, especially when it becomes sufficient to trigger the formation of planetesimals in the trap (see Sect.~\ref{sec:results_planetesimalformation}). In the most blocking conditions ($\alpha_\mathrm{visc} = 10^{-3}$, $\delta_\mathrm{turb} = 10^{-4}$), traps carved by a $300 \mathrm{~M_\oplus}$ planet at $10$ au (or a $100 \mathrm{~M_\oplus}$ planet at $40$ au) leads to pebble flux up to two orders of magnitudes lower compared to permeable traps ($F_\mathrm{peb}(r_\mathrm{ice}) \approx 10^{-7} - 10^{-6} \mathrm{~M_\oplus~yr^{-1}}$). In our parameter study, the most efficient reduction of the pebble flux (and hence of the delivery of volatiles) is reached for resistant grains ($v_\mathrm{frag} = 3 \mathrm{~m~s^{-1}}$) along with low viscosity and weak turbulence, together allowing for the formation of pebbles with Stokes number close to $0.1$, and resulting in $F_\mathrm{peb}(r_\mathrm{ice}) \lesssim 10^{-7} \mathrm{~M_\oplus~yr^{-1}}$.

Comparing with the water vapour reservoir observed with JWST, we see that fully permeable dust traps are close to but slightly below the water reservoir measured in smooth discs ($M_\mathrm{H_2O}^\mathrm{cold} \approx 1-10 \mathrm{~\mu M_\oplus}$). Increasing the initial disc mass ($M_\mathrm{disc} = 0.15 \mathrm{~M_\odot}$, see lower-right panel in Fig.~\ref{fig:FigY_Mvap_Fpeb}) yields a higher pebble flux closer to measurements, though decreasing $\eta$ (see Eq.~\ref{eq:F_peb_to_M_cold}) would have a similar effect. We find a good agreement between JWST measurements and our simulations with highly to moderate trap leakiness ($M_\mathrm{H_2O}^\mathrm{cold} \approx 0.1-1 \mathrm{~\mu M_\oplus}$). However, large uncertainties remain when converting cold water vapour mass in pebble flux, such that it is challenging to draw firm conclusions. 

Considering CI Tau as a lower limit\footnote{The disc SR 4 has a lower cold water vapour mass than CI Tau but larger error bars, such that we pick CI Tau as a lower limit.} for the cold water vapour mass that can be detected with JWST ($M_\mathrm{H_2O}^\mathrm{cold} \approx 10^{-1} \mathrm{~\mu M_\oplus}$), we can compute the amount of dust traps in our parameter grid that have sufficient pebble flux to yield a detectable cold water signal. For $\eta = 10^{3}$, and focusing on the pebble flux at 1 Myr, the fraction of discs hosting dust traps leaky enough to yield a detectable signal is $67 \%$. This increases to $82 \%$ for a lower $\eta = 10^{2}$, but decreases to only $25 \%$ of our sample for $\eta = 10^{4}$. The fact most traps yield a detectable signal agrees with the fact most observed discs with JWST hosting deep or shallow dust traps resolved by ALMA display a cold water vapour excess \citep{banzatti2023JWST, banzatti2025water, arulanantham2025jdisc, gasman2025minds}. This also highlights that the analysis of cold water vapour lines and their variations can be a reliable method to quantify the leakiness of dust traps \citep{krijt2025cosmic}, unlike the C/O ratio which we discuss further in Sect.~\ref{sec:discussion_JWST_CO}.

\section{Discussion} 
\label{sec:discussion}

\subsection{The different leaking regimes}
\label{sec:discussion_leaking_regime}

\subsubsection{Leaking regimes identified in this work}

We have seen in this work that the values of some key parameters ($\alpha_\mathrm{visc}$, $\delta_\mathrm{turb}$, $M_\mathrm{p}$ and $a_\mathrm{p}$) have a great impact on the leaking of dust traps, giving rise to a variety of leaking regimes impacting the amount of outer disc dust and icy volatiles received by the inner disc. We have identified 4 regimes which we discuss in this Section. 

\paragraph{Highly (fully) permeable ($\mathcal{B}\sim$  $0$ -- $0.2$):} We identified some dust traps as being \textit{highly (fully) permeable}. Their blocking efficiency is typically $\mathcal{B}\sim$  $0$ -- $0.2$, indicating that the outer disc material is transported inward nearly identically to a smooth disc. In this case, dust particles of any size pass through the trap. Dust traps may be highly permeable due to (1) high global viscosity ($\alpha_\mathrm{visc} \gtrsim 3 \times 10^{-3}$) weakening the gap depth and giving a stronger inward velocity component to dust particles, (2) strong local turbulence ($\delta_\mathrm{turb} \gtrsim 3 \times 10^{-3}$) boosting diffusion and lowering the maximum particle size, (3) low planet mass ($M_\mathrm{p} \lesssim 100 \mathrm{~M_\oplus}$) decreasing the local pressure maximum hence the outward dust drift component, and (4) close distance from the star ($a_\mathrm{p} \lesssim 5 \mathrm{~au}$) modifying the dust properties.

\paragraph{Super-leaky traps ($\mathcal{B}<0$):} Some highly permeable traps turned out to have negative blocking efficiency $\mathcal{B} < 0$, which we labelled as being \textit{super-leaky}. In that case, the full permeability of the dust trap coupled with the much higher gas velocity in the gap centre boosts the inward transport of solids. This results in a higher delivery of solids in the planet-hosting disc compared to a smooth disc, though not by much, as the blocking efficiency remains close to zero. This regime is favoured when the gap is particularly deep and wide, boosting the inward velocity by a greater factor over a longer duration. This happens for massive Jupiter-mass planets ($M_\mathrm{p} \gtrsim 300 \mathrm{~M_\oplus}$) and/or planets located closer to the star ($a_\mathrm{p} \lesssim 5 \mathrm{~au}$).

\paragraph{Moderately leaky ($\mathcal{B}\sim$ $0.2$ -- $0.9$):} These dust traps are strong enough to block the largest pebbles. Even though these particles contain most of the solid mass, their continuous fragmentation into smaller grains allows for a significant fraction of the outer disc mass to leak through the trap. We denote these traps as \textit{moderately leaky} as often called in the literature \citep[e.g.,][]{stammler023leaky}. Their blocking efficiency ranges between $\mathcal{B}\sim$ $0.2$ -- $0.9$. The inner part of discs with a leaky trap typically receives a smaller pebble flux, reduced by a few factors compared to an equivalent smooth disc ($F_\mathrm{peb}(1 \mathrm{~au}) \approx 10^{-5} \mathrm{~M_\oplus~yr^{-1}}$, Fig.~\ref{fig:FigY_Mvap_Fpeb}). Because the pebble flux is slowed down, it is also longer-lived. Overall, moderately leaky traps will result in a long-lived oxygen-rich inner disc, with a flux of icy pebbles sufficient to result in detectable cold water vapor lines \citep{banzatti2025water}.

\paragraph{Highly blocking ($\mathcal{B} > 0.9$):} Some dust traps are strong enough to only allow the smallest grains (close to monomer size) to leak through. In that case, the dust evolution is slowed down importantly, resulting in \textit{highly blocking} traps with blocking efficiency $\mathcal{B} > 0.9$ over the disc lifetime (5-10 Myr). Most of the dust remains in the dust trap, and only a small pebble flux reaches the inner disc, which may result in a significant variation of the inner disc composition. Compared with simpler dust evolution models (Sect.~\ref{sec:results_2pop_vs_DustPy}), our parameter study with full dust evolution simulations show that dust traps are leakier than previously thought, such that highly blocking dust traps require specific conditions. Typically, they can be found for the smallest pair of viscosity and turbulence we explored ($\alpha_\mathrm{visc} = 10^{-3}$, $\delta_\mathrm{turb} = 10^{-4}$) and for planet mass $M_\mathrm{p} \gtrsim 100 \mathrm{~M_\oplus}$ or for more resistant grains. Moreover, highly blocking traps are typically made possible due to the efficient formation of planetesimals, sequestrating icy pebbles and hence reducing the mass flux through the trap (see Fig.~\ref{fig:FigX_Mpla_end}). However, if planetesimal formation is inactive, e.g., due to planet-induced turbulence \citep{binkert2023three}, these dust traps could fall back on the moderately leaky regime. Interestingly, the role of planetesimal formation in producing highly blocking dust traps was not found in previous works using simpler dust evolution models \citep{kalyaan2023effect}; this highlights further the need to use full dust evolution simulations when simulating dust traps.

\subsubsection{Can a perfectly blocking dust trap exist?}
\label{sec:discussion_perfect_blocking}

A regime that we do not identify in our parameter study is the case of a \textit{perfectly blocking} dust traps, in which case the pebble flux across the trap is zero. Such trap could technically be found if monomers themselves are blocked in the dust trap ($\mathrm{St_{crit}} < \mathrm{St_{mon}}$), as was found by \citet{pinilla2024survival} when modelling PDS 70 to explain the observation of water vapour in its inner disc \citep{perotti2023water}. However, the size of monomers is not well constrained \citep[though see][]{tazaki2022size, hensley2023astrodust}. Monomers in reality would also likely have a certain size distribution \citep{dominik1997physics, xiang2020detailed}, or be breakable by collisional fragmentation. It is thus challenging to set clear predictions on the existence of fully blocking traps. We speculate whether deep transition discs for which we do not measure a dusty inner disc \citep{rota2024correlation, rota2025correlation} may be examples of disc conditions where traps are perfectly blocking.

\subsection{Leaky traps and inner disc composition with JWST}
\label{sec:discussion_JWST}

\subsubsection{Long-lived O-rich inner disc}
\label{sec:discussion_JWST_Oxygen}

With its increased sensitivity and spectral resolution, JWST has allowed us to dive deeper than ever into the chemical reservoir of the inner part ($\sim$$5 \mathrm{~au}$) of protoplanetary discs. As mentioned in Sect.~\ref{sec:intro}, the first years of science with JWST has confirmed early trends seen with Spitzer, namely that the inner disc exhibits a significant compositional diversity. The observed masses of key species (e.g., $\mathrm{H_2O}$, $\mathrm{CO_2}$, or $\mathrm{C_2H_2}$) varies by orders of magnitude across observed protoplanetary discs \citep{arulanantham2025jdisc}. Moreover, some molecules are so depleted in some discs that they are not even detected in spectra, such as water in some discs around very low-mass stars \citep{tabone2023rich, arabhavi2025minds}. Though several mechanisms have been suggested to explain this diversity, the presence of a dust trap (i.e., local pressure maximum) efficiently blocking icy pebbles prior to sublimation is perhaps the most popular \citep{tabone2023rich, vandishoeck2024probing, arabhavi2026molecular}.

Earlier dust transport models designed to study the impact of outer dust traps on the inner disc composition were often based on the two-population algorithm, and found that they can have a significant importance \citep{kalyaan2023effect, mah2024mind, sellek2025co2}. In fact, \citet{mah2024mind} argues that most outer dust traps result in a depletion of oxygen in the inner disc \citep[Regime III, see figure 1 in][]{mah2024mind}, as water (the main carrier of oxygen) remains in the ice mantle of pebbles halted in the trap. 

However, as we have discussed in Sect.~\ref{sec:results_2pop_vs_DustPy}, models based on \texttt{two-pop-py} can underestimate by several orders of magnitude the amount of icy pebbles making it through the dust trap. From our parameter study based on a full dust evolution model, we thus find that dust traps are much leakier than previously thought, and across a broader parameter space. As a result, we argue that the presence of outer disc traps will most likely result in a less intense but longer-lived oxygen-rich inner disc, as icy pebbles slowly leak inward. It is still possible for dust traps to be strong enough to deplete the inner disc in oxygen, but that requires specific conditions: the viscosity and turbulence must be weak, and planetesimals have to form efficiently in the trap. But in that case, it is planetesimal formation rather than the dust trap itself that is capable to significantly alter the inner disc composition (e.g., oxygen depletion).

Our conclusion aligns with the wealth of data that has been gathered by JWST demonstrating that the cold infrared lines of water vapour, whose detection requires an ongoing and non-negligible flux of icy pebbles \citep{vlasblom2025understanding}, are often detected in discs that host shallow or even deep dust traps \citep{banzatti2023JWST, banzatti2025water, arulanantham2025jdisc, gasman2025minds}.

\subsubsection{Inner disc C/O ratio}
\label{sec:discussion_JWST_CO}

Previous works have argued that the presence of an outer dust trap may result in gas-phase $\mathrm{C/O} > 1$ in the inner disc \citep[e.g.,][]{mah2024mind}. The reason is that if icy pebbles are efficiently trapped in the cold outer region, the delivery of oxygen (in the form of water) to the inner disc is halted. Meanwhile, a fraction of the carbon reservoir is stored in highly-volatile components (e.g., CH$_4$) which sublimates into the gas-phase in the far outer regions. As the inward gas flow is not perturbed by the dust trap, it moves inward towards the inner disc, resulting in gas-phase C/O ratio above unity.

However, as our parameter study finds that it is more challenging than previously thought to halt the delivery of oxygen-rich icy pebbles to the inner disc, the question is: is the leaking of icy pebbles through the dust trap higher than the flow of carbon-rich gas? Comparing these two quantities can effectively tell us whether dust traps can result in inner disc gas-phase C/O ratio above or below unity. With solar carbon abundances \citep[$\mathrm{C/H} = 2.69 \times 10^{-4}$, see][]{asplund2009chemical}, and assuming an upper limit that $10\%$ of carbon is stored in highly-volatile CH$_4$ molecules capable of elevating the C/O ratio \citep{mah2023close}, we can estimate the flux of C-rich gas\footnote{The volatile reservoir of carbon also consists of CO and CO$_2$, but we do not consider them as they contain oxygen and hence are not able to elevate the gas-phase $\mathrm{C/O}$ above unity.} as $F_\mathrm{C-bearing} =  \frac{(\mathrm{C/H})}{10} F_\mathrm{gas}$, where $F_\mathrm{gas}$ is the gas flux from our simulations. Meanwhile, the flux of oxygen is $F_\mathrm{O-bearing} = f_\mathrm{H_2O,ice}F_\mathrm{peb}$, assuming a lower limit for the mass fraction of water ice on pebbles to $f_\mathrm{H_2O,ice} = 0.2$ (Sect.~\ref{sec:results_M_H2O}).

\begin{figure}
    \centering
    \includegraphics[width=\columnwidth]{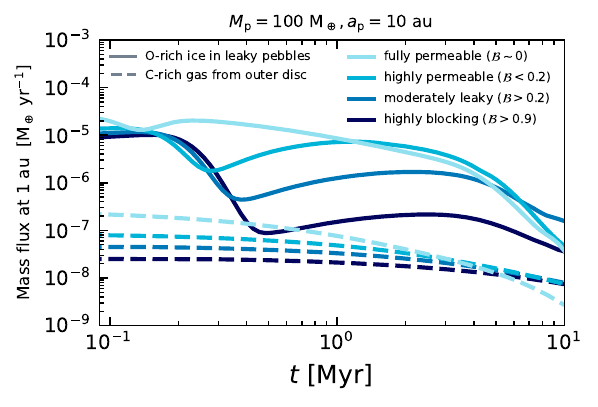}
    \caption{Mass flux of oxygen (stored in water ice, solid lines) and carbon (in gas-phase CH$_4$, dashed lines) at 1 au as a function of time, for a disc with a gap carved by a $100 \mathrm{~M_\oplus}$ planet at $10 \mathrm{~au}$. As in Fig.~\ref{fig:FigY_Mvap_Fpeb}, we focus on simulations with $\delta_\mathrm{turb} = 10^{-4}$ and varying viscosity, as these sample well the different leakiness regimes. We see that the pebble flux is always above the flux of C-rich gas, meaning that it is challenging for outer dust traps to result in inner disc gas-phase C/O above unity.}
    \label{fig:Fig_CO_ratio}
\end{figure}

We present in Fig.~\ref{fig:Fig_CO_ratio} the flux of oxygen (in the form of icy pebbles) and carbon (in the form of gas-phase CH$_4$) reaching the iceline (1 au), for our fiducial model of a trap carved by a $100 \mathrm{~M_\oplus}$ at $10 \mathrm{~au}$ (upper-left panel in Fig.~\ref{fig:FigY_Mvap_Fpeb}). As we did earlier, we focus on models with $\delta_\mathrm{turb} = 10^{-4}$ and varying viscosity $\alpha_\mathrm{visc}$, as these sample well the different blocking efficiency regimes (Fig.~\ref{fig:FigX_Blockiness_ap_Mp}). We see that even for highly blocking dust trap, the flux of oxygen is orders of magnitude higher than the flux of carbon. Upon checking with our full sample ($250$ different dust traps), $4$ dust traps\footnote{With a higher water ice mass fraction on pebbles $f_\mathrm{H_2O, ice} = 0.5$ \citep{lodders2003solar}, this number falls to $1$ dust trap, corresponding to a $300 \mathrm{~M_\oplus}$ planet at $40 \mathrm{~au}$.} manage to halt icy pebbles sufficiently for carbon to dominate and increase the gas-phase C/O ratio above unity, which are those where $\mathcal{B} > 0.99$ and planetesimals form efficiently. Similarly to our conclusion in Sect.~\ref{sec:discussion_JWST_Oxygen}, it is thus the efficient formation of ice-rich planetesimals more than dust traps themselves that manages to raise the C/O ratio in the inner disc.

As a growing sample of protoplanetary discs observed with JWST, in particular around very low-mass stars, exhibit inner disc gas-phase $\mathrm{C/O} > 1$ \citep[][]{tabone2023rich, arabhavi2024abundant, kanwar2024minds, long2025first, colmenares2024jwst}, our findings raise a question: if achieving inner disc $\mathrm{C/O} > 1$ through the presence of outer disc dust traps requires them to be highly blocking, along with efficiently forming planetesimals, does this mean that such traps are common in observed discs around very low-mass stars? Instead, other processes may be contributing to the inner disc carbon budget, such as the thermal or chemical processing of carbon stored in refractory species \citep{houge2025burned}, the chemical transformation of highly-volatile CO into CH$_4$ \citep{sellek2025chemical}, the presence of inner disc cavities \citep{mallaney2026protoplanetary}, or the vertical distribution of dust grains \citep{grant2025minds}. In that case, we suggest that the cold water vapour emission may be a better observational tracer than the gas-phase C/O ratio to quantify the leakiness of dust traps, though it may require further coupling between dust evolution models and thermochemical models \citep{vlasblom2025understanding} to produce spectra directly comparable to observations. This is also supported by the fact that most dust traps in our parameter grid yield a detectable cold water vapour mass, considering CI Tau as a lower limit (Sect.~\ref{sec:results_M_H2O}).

In conclusion, mapping the parameter space of dust traps with full dust evolution models suggests that dust traps are more likely to have a limited impact on the inner disc composition, maintaining a long-lived oxygen-rich inner disc (hence low inner disc C/O ratio), with some variation in species abundances on a population level that agrees with observations \citep{banzatti2023JWST, banzatti2025water, arulanantham2025jdisc, gasman2025minds, krijt2025cosmic}. Still, under the right conditions (low turbulence, low viscosity, efficient planetesimal formation), highly blocking traps can slow down icy pebbles by orders of magnitude and deeply affect the inner disc composition. 

\subsubsection{The impact of inner vs. outer traps}
\label{sec:discussion_JWST_inner}

\begin{figure}
    \centering
    \includegraphics[width=\columnwidth]{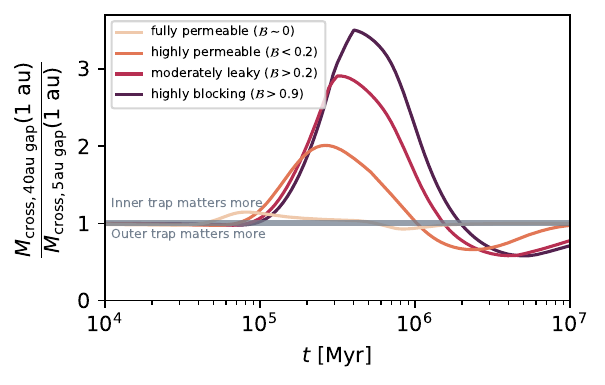}
    \caption{Ratio of the cumulative mass flux (Eq.~\ref{eq:M_crossed}) reaching 1 au of simulations with outer traps at 40 au relative to simulations with inner traps at 5 au both carved by a $100 \mathrm{~M_\oplus}$ planet, for different blocking efficiency regime as in Fig.~\ref{fig:FigY_Mvap_Fpeb}. The ratio changes from above to below unity, indicating that in early stages ($t < 1 \mathrm{~Myr}$) inner dust traps have a greater impact on the delivery of material to the iceline, while at later stages ($t > 1 \mathrm{~Myr}$) their overall higher leakiness makes them less impactful.}
    \label{fig:Fig_ratio_inner_outer}
\end{figure}

From Fig.~\ref{fig:FigX_Blockiness_ap_Mp}, we see that highly blocking dust traps are more easily produced in the outer disc, due to how dust properties change with distance from the star. However, as noted by \citet{kalyaan2023effect}, even if inner dust traps are leakier, a larger fraction of the total disc reservoir has to go through the trap. Over long timescales, this could compensate for the higher leakiness of inner traps, and make them more effective at controlling the inner disc composition than outer traps. To investigate the long term effect of inner vs. outer traps. we can instead compare the cumulative mass flux $M_\mathrm{cross}$ through $1 \mathrm{~au}$ (see Eq.~\ref{eq:M_crossed}) for simulations with traps at $40 \mathrm{~au}$ and $5 \mathrm{~au}$. We present their ratio in Fig.~\ref{fig:Fig_ratio_inner_outer} for identical disc models, that is assuming both traps are carved by a $100 \mathrm{~M_\oplus}$ planet in a disc with identical viscosity and turbulence. We show the results of different simulations varying the viscosity leading to varying trap blocking efficiency , as in Fig.~\ref{fig:FigY_Mvap_Fpeb}. We see that the ratio is above unity for $t < 1 \mathrm{~Myr}$, indicating that the inner part of discs with outer dust traps receive more material at early times. This is because the pebble flux reaching $1 \mathrm{~au}$ takes longer than 1 Myr to decrease because of the effect of a trap at 40 au. For $t > 1 \mathrm{~Myr}$, the ratio is below unity, indicating that on the long run, discs with inner traps ($r \sim 5 \mathrm{~au}$) deliver more material to the inner disc. 

Overall, for traps carved by an identical planet mass, inner dust traps are the most effective at early times ($\lesssim$$1 \mathrm{~Myr}$) in regulating the inner water delivery \citep{kalyaan2021linking, kalyaan2023effect, easterwood2024water, krijt2025cosmic}, while at later times ($\gtrsim$$1 \mathrm{~Myr}$), outer dust traps become the most effective. This behaviour was not identified by \citet{krijt2025cosmic}, which we explain by the fact they assumed a trap blocking efficiency constant with the trap position, while we find inner traps to be leakier (Fig.~\ref{fig:FigX_Blockiness_ap_Mp}). Further investigating the impact of inner vs. outer traps could be an interesting avenue for future works, especially implementing several traps per disc in \texttt{DustPy} simulations.

\subsection{Planetesimal formation and pebble accretion beyond $M_\mathrm{iso}$}

Apart from impacting how the disc composition evolves, the leakiness of dust traps has an impact on the onset of planet formation itself. We have seen in Sect.~\ref{sec:results_planetesimalformation} that leakiness may prevent the dust trap from reaching the high dust densities required to trigger planetesimal formation. The onset of the streaming instability in the dust trap is thus limited to a subset of the parameter space, associated with weak local turbulence ($\delta_\mathrm{turb} \lesssim 10^{-4}$) and a deep gap as reached in our model with low viscosity ($\alpha_\mathrm{visc} \lesssim 3 \times 10^{-3}$) and high planet mass ($M_\mathrm{p} \geq 100 \mathrm{~M_\oplus}$). Naturally, the onset of planetesimal formation in dust traps depends on a more extensive parameter space than the one explored in this work, such as the initial dust-to-gas ratio \citep[e.g.,][]{andama2024which}. Still, modelling self-consistently the leaking process is a necessary ingredient to fully explore what may or not happen in dust traps \citep{lyra2009planet}. 

Dust traps also impact the later stages of planet formation, namely the growth of planetary embryos through pebble accretion. Pebble accretion is generally assumed to stop once the planetary embryo reach the pebble isolation mass $M_\mathrm{iso}$ \citep[e.g.,][]{lambrechts2014separating, ataiee2018how, bitsch2018pebble}. Beyond that mass, which we estimate in our disc model to $M_\mathrm{iso} \sim 15 \mathrm{~M_\oplus}$ at $10 \mathrm{~au}$ following \citet{johansen2017forming}, the protoplanet deeply modifies the local gas profile, and inward-drifting pebbles become trapped in the local pressure maximum. However, we have seen in this work that the fragmentation of trapped pebbles leads to a constant replenishment of small grains able to leak (even for $M_\mathrm{p} > 20 ~M_\mathrm{iso}$). Depending on the trajectory of small grains once in the planetary gap, a fraction of the leaky flux may thus be accreted by the protoplanet, even beyond $M_\mathrm{p} > M_\mathrm{iso}$ \citep[see also discussion in][]{drazkowska2019including}. Using 3D hydrodynamic simulations with a fixed particle size, \citet{vanclepper20253D} finds that a relatively small fraction of small grains may be accreted by the planet. Considering small grains may re-coagulate in the planetary gap to larger Stokes number (Fig.~\ref{fig:Fig2_SigmaD_gap_nogap}), the accreted fraction may be larger. Concerning the effect on other planets that may form interior to a dust trap, permeable or moderately leaky dust traps would have a limited impact, but could eventually modify the system architecture, favouring terrestrial planets or super-Earths for low pebble fluxes \citep{lambrechts2019formation, johansen2021pebble}. On the other hand, highly blocking traps could prevent the growth of inner planets, while simultaneously converting the trapped mass into planetesimals in the outer system. That is, of course, providing that the outer planet forms first, which is not expected from the perspective of faster growth by pebble accretion in the inner disc \citep{lau2024sequential, danti2025super}.

\subsection{Limitations}
\label{sec:discussion_limitation}

In this work, we have performed an extensive parameter space exploration of leaky dust traps over the whole disc evolution (with $t_\mathrm{end} = 10 \mathrm{~Myr}$). Our focus has been on understanding how the specificity of dust coagulation, fragmentation, and transport, can alter the leakiness of traps and hence the delivery of outer disc material to the inner disc. For this we used the state-of-the-art code \texttt{DustPy}, simulating dust evolution with a full grain size distribution over a one-dimensional radial grid. Even though this represents a significant improvement compared to simpler dust evolution models, there are still many processes that cannot be captured accurately by \texttt{DustPy}, notably due to its one-dimensional nature.

When using sophisticated 2D or 3D hydrodynamic simulations, it is possible to simulate self-consistently the gap-opening process \citep[e.g.,][]{kanagawa2017modelling, weber2018characterising, weber2019predicting, haugbolle2019probing, duffell2020empirically}. This gives access to accurate gap depth and shape, along with the gas flow patterns in and around the planetary gap \citep{petrovic2024material, huang2025leaky, vanclepper20253D, vanclepper2025DRIVE}. Though our gap prescription is based on 2D hydrodynamic simulations performed by \citet{kanagawa2017modelling}, some differences may arise depending on the parameter space that is explored \citep{pfeil2025fragmentation}. Note, however, that hydrodynamic simulations are restricted to a single particle size or a few non-interacting size bins (i.e., no mass exchange). Their improvements in modelling the gap and gas flow patterns is thus counterbalanced by the fact they cannot capture well the coagulation and fragmentation of dust grains, which we have seen in this work is important for the modelling of dust trap leakiness. Moreover, hydrodynamic simulations are limited in their temporal span ($t_\mathrm{end} \ll 1 \mathrm{~Myr}$), and hence they cannot evolve the transport of dust over the whole disc lifetime. Because leaking slows down dust evolution, simulating dust transport over long timescales, as can be done with \texttt{DustPy}, gives the possibility to perform a long-term analysis of the pebble flux. Overall, using either a full grain size distribution or a multi-dimensional hydrodynamic simulation, both have their advantages and limitations. Though there has been some efforts in combining these two approaches \citep{drazkowska2019including}, there is still work to do to build a new framework including both effects. Nevertheless, the recent release of \texttt{TriPod} \citep{pfeil2024tripod, kaufmann2025tripod}, the upgrade of the \texttt{two-pop-py} algorithm, opens new exciting doors for a better simple treatment of dust evolution in hydrodynamic simulations \citep{pfeil2025fragmentation}. 

%mention NC/CC???? 
%Focusing on the Solar System, Stammler+2023 used such an approach to demonstrate that the dust trap induced by the potential early formation of a Jupiter-sized planet would not be sufficient to prevent dust particles to move through the planetary gap, overall contaminating the inner disc in outer disc material. Our knowledge of the leakiness of dust traps thus completely changes how we conceptualize the formation of the Solar System to make sense of meteoritic data, favouring alternative models. Notably, Lichtenberg+2021 showed that a temporal separation between the formation of NC and CC parent bodies could explain meteoritic measurements without requiring a radial separation by Jupiter. Moreover, Nerea+2025 showed that if the CC material is stored in chondrule, which have much higher resistance to fragmentation, a dust trap could still segregate material.
% Also Colmenares paper

\section{Conclusions} 
\label{sec:conclusions}

In this paper, we performed a large-scale parameter space exploration of dust transport in protoplanetary discs containing a dust trap, assumed to have a planetary origin, with the state-of-the-art dust evolution code \texttt{DustPy} \citep{stammler2022dustpy}. Our goal was to quantify the leakiness of different dust traps over the disc lifetime and map out in which area of the parameter space different leakiness regimes can be found. We then discussed the impact of our results on the delivery of volatiles to the inner part of discs, and how it may relate to the chemical diversity observed with JWST. Our findings are summarized as follows:

\begin{itemize}
    \item A planet-induced gap results in the establishment of a local pressure maximum outside the gap. The pressure maximum acts as a dust trap in which pebbles, whose motion follows the opposite direction of the pressure gradient, accumulate. Meanwhile, small grains are well coupled to the gas and leak through the dust trap (Fig.~\ref{fig:dust_velocity_vs_Stokes}). Even though small grains make up a minor fraction of the total solid mass, the continuous fragmentation of pebbles gradually transfers mass into leaking grains (Fig.~\ref{fig:Fig2_SigmaD_gap_nogap}). Over time, most of the outer disc solid reservoir may flow through the gap and contaminate the inner disc (Fig.~\ref{fig:F_peb_blockiness_onesim}). Including self-consistently the mass exchanges between pebbles and grains is a fundamental ingredient in quantifying the leakiness of trap.

    \item We perform a large parameter space exploration, investigating the impact of the disc viscosity $\alpha_\mathrm{visc}$, the disc turbulence $\delta_\mathrm{turb}$, the planet mass $M_\mathrm{p}$ and location $a_\mathrm{p}$, and the fragmentation velocity $v_\mathrm{frag}$. At a fixed location, the dust trap carved by a Jupiter-mass planet is much less leaky than that of a smaller planet (Fig.~\ref{fig:FigX_Blockiness_ap_Mp}). Furthermore, dust traps in the inner disc ($r \leq 5 \mathrm{~au}$) are leakier than traps in the outer disc ($r \geq 40 \mathrm{~au}$, see Fig.~\ref{fig:FigX_Blockiness_ap_Mp}) due to the change in dust properties in the inner vs. outer disc (e.g., temperature, Stokes number). Therefore, while at early times ($t \lesssim 1 \mathrm{~Myr}$) inner traps are the most effective in reducing the pebble flux, outer traps become more important at later times ($t \gtrsim 1 \mathrm{~Myr}$) as they are less leaky (Sect.~\ref{sec:discussion_JWST_inner}). More resistant dust grains ($v_\mathrm{frag} = 3 \mathrm{~m~s^{-1}}$) and weak turbulence ($\delta_\mathrm{turb} \leq 3 \times 10^{-4}$) yield much more blocking dust traps (Fig.~\ref{fig:FigX_Blockiness_vf3}), though the resulting pebble size is larger than what is measured in discs with ALMA \citep[e.g.,][]{jiang2024grain, tong2026turbulence, luo2026almasurveygasevolution}.

    \item From our parameter study, we find most dust traps to be highly permeable or moderately leaky (Fig.~\ref{fig:FigX_Blockiness_ap_Mp}), in which case the pebble flux is only reduced by up to a factor of a few and a large fraction of the outer disc reservoir reaches the inner disc over the disc lifetime (5 Myr). As a result, most dust traps will result in a long-lived oxygen-rich inner disc (Sect.~\ref{sec:discussion_JWST_Oxygen}), paired with gas-phase C/O ratio below unity (Sect.~\ref{sec:discussion_JWST_CO}). This provides a possible explanation for the cold water excess commonly detected in discs with JWST, including those hosting shallow or deep traps \citep{banzatti2023JWST, banzatti2025water, gasman2025minds, temmink2025minds, krijt2025cosmic}. To relate dust evolution models directly to measured cold water excess requires more work (Sect.~\ref{sec:results_M_H2O}), notably on constraining the observable column $\eta$ or coupling dust simulations with thermochemical models.

    \item We compared our results with that of simpler dust evolution models based on the two-population algorithm \citep[\texttt{two-pop-py}, see][]{birnstiel2012simple}, used in earlier works to study the impact of outer dust traps on the inner disc composition. We find that simpler dust evolution models may overestimate the amount of dust blocked in the trap sometimes by several orders of magnitude (Sect.~\ref{sec:results_2pop_vs_DustPy}). As a result, we find that dust traps are leakier than previously thought, such that the inner and outer disc reservoirs are well connected over a broader parameter space.

    \item In the right conditions (low viscosity and weak turbulence), highly blocking dust traps reducing the pebble flux by orders of magnitude can be produced (Fig.~\ref{fig:FigX_Blockiness_ap_Mp}). The reduction in pebble flux is such that highly blocking dust traps are capable of deeply affecting the inner disc composition, and may result in an oxygen-poor inner disc with gas-phase $\mathrm{C/O} > 1$ (Sect.~\ref{sec:discussion_JWST_CO}). However, the existence of highly blocking traps typically rely on the efficient formation of planetesimals taking place in the trap. If planetesimals did not form (Fig.~\ref{fig:Fig_Fleak_with_without_planetesimals}), or if they undergo ablation over time \citep{eriksson2021fate}, the traps would be much leakier. In that case, it means that it is planetesimal formation, rather than dust traps themselves, which are capable of significantly altering the inner disc composition. Moreover, within the parameter space we explore, we do not find any fully blocking dust trap (Sect.~\ref{sec:discussion_perfect_blocking}).

    \item Depending on the disc viscosity and gap depth, we find some discs with highly permeable traps to have a more efficient dust transport than in an equivalent smooth disc. We refer to these traps as \textit{super-leaky}, and this extreme leakiness can occur because of the higher inward gas velocity during the crossing of the gap.
   
\end{itemize}

In the end, the leakiness of dust traps is not a simple matter of finding the critical size of grains that are experiencing leaking in contrast to filtration, but requires complex coagulation and fragmentation computation based on a full grain size distribution. 2D and 3D hydrodynamic simulations offer great opportunities to probe complementary aspects of dust trap leakiness (gas flow patterns, gap-opening process, pebble accretion beyond $M_\mathrm{iso}$), but do not give a complete picture in terms of size distribution and long-term temporal evolution. In the future, it is thus essential to try and bring the strength of both methods together, such as pioneered by \citet{drazkowska2019including} and \cite{pinilla2024survival}. In that matter, new dust evolution models developed for implementation into hydrodynamical codes, such as \texttt{TriPoD}, offer promising avenues \citep{pfeil2025fragmentation}, though such approaches are yet to be tested \citep[see][]{eriksson2026turning}.

%--------------------------------------------------------------------
\begin{acknowledgements} 

A.H. thanks Sebastiaan Krijt, Joe Williams and Claudia Danti for insightful discussions. A.H. and A.J. acknowledge funding from the Carlsberg Foundation (Semper Ardens: Advance grant FIRSTATMO).

\end{acknowledgements}

\bibliographystyle{aa}
\bibliography{paper}

\begin{appendix}

\section{Removing the dust reservoir}
\label{sec:appendix_removingdust}

\begin{figure}
    \centering
    \includegraphics[width=\columnwidth]{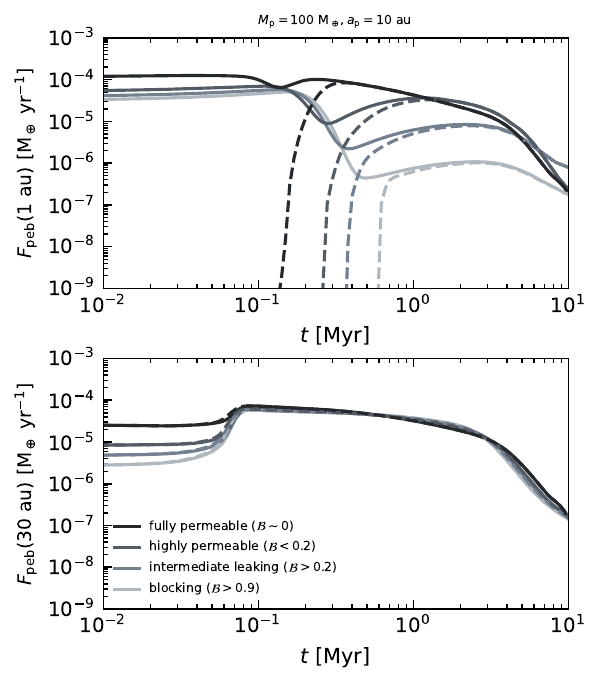}
    \caption{Pebble flux $F_\mathrm{peb}$ inside the planetary gap ($1 \mathrm{~au}$, upper panel) and outside the planetary gap ($30 \mathrm{~au}$, lower panel) for a disc with a $100 \mathrm{~M_\oplus}$ planet at $10 \mathrm{~au}$. Solid lines correspond to full simulations, while dashed lines show those where dust has been initially emptied out inside of the gap. The different shades indicate different simulations with $\delta_\mathrm{turb} = 10^{-4}$ and varying viscosity, which sample well the different blocking efficiency regime. Low viscosity ($\alpha_\mathrm{visc} = 10^{-3}$) is highly blocking, while strong viscosity ($\alpha_\mathrm{visc} = 10^{-2}$) are highly permeable.}
    \label{fig:F_peb_appendix}
\end{figure}

As mentioned in Sect.~\ref{sec:methods}, we initially empty the dust reservoir inside the planetary gap, such that with ongoing evolution, the dust surface density inside the gap has to originate from the leakiness of the dust trap. Moreover, this reduces the computational expense of \texttt{DustPy}, highly simplifying our parameter space exploration. To ensure that this approach does not modify our results (pebble flux, blocking efficiency ), we compare the results of simulations with inner disc dust emptied out with full simulations, which were shown in Fig.~\ref{fig:FigY_Mvap_Fpeb}. The different simulations correspond to $\delta_\mathrm{turb} = 10^{-4}$ and a sample of viscosity $\alpha_\mathrm{visc}$, which samples different blocking efficiency regime.

We present in Fig.~\ref{fig:F_peb_appendix} the pebble flux at 1 au and 30 au with and without inner disc dust initially present for a disc with a $100 \mathrm{~M_\oplus}$ planet at $10 \mathrm{~au}$. When inner disc dust is initially emptied out, we see that the pebble flux reaching 1 au takes longer to increase, which corresponds to the reservoir of solids inside the gap that was removed. Once the pebble front coming from outside the gap reaches 1 au, the pebble flux in both simulations becomes similar. The timescale for the pebble flux of both simulations to adjust can inform us on the time it takes for the inner disc to feel the effect of the outer dust trap, referred to as $t_\mathrm{delay}$ in \citet{krijt2025cosmic}. We see that it is not necessarily identical to the drift timescale, but that the blocking efficiency of the dust trap slows down the transport of particles and increase that delay. This is due to the cycle of fragmentation and re-coagulation dust grains have to go through in order to escape the dust trap. Focusing on the blocking efficiency after 5 Myr, we see that this quantity is unaffected by our approach.

Overall, emptying the inner dust disc is a powerful way to reduce the computational expense of \texttt{DustPy} and allows to directly relate the dust mass of the inner disc to what has leaked through the trap. Otherwise, the implementation of a passive tracer on dust particles \citep[as in][]{pfeil2025fragmentation} would be necessary to disentangle what fraction of the inner disc mass originates from the inner or outer disc.

\section{Pebble flux in \texttt{DustPy} vs. \texttt{two-pop-py}}
\label{sec:appendix_twopop_DustPy}

\begin{figure}
    \centering
    \includegraphics[width=\columnwidth]{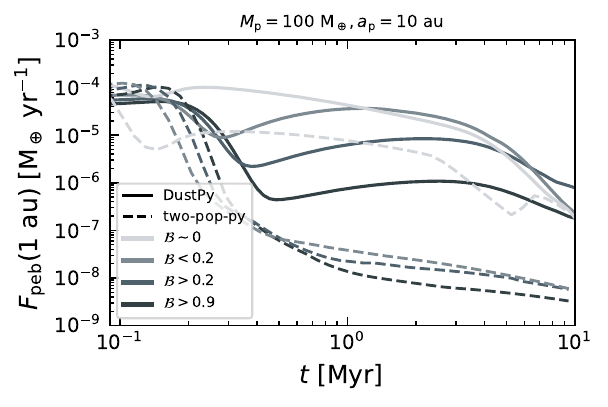}
    \caption{Pebble flux $F_\mathrm{peb}$ inside the planetary gap ($1 \mathrm{~au}$) for a disc with a $100 \mathrm{~M_\oplus}$ planet at $10 \mathrm{~au}$. Solid lines correspond to full simulations with \texttt{DustPy}, while dashed lines show the results of \texttt{two-pop-py}. Different shades indicate different simulations with $\delta_\mathrm{turb} = 10^{-4}$ and varying viscosity (see also Fig.~\ref{fig:F_peb_appendix}).}
    \label{fig:F_peb_twopop_dustpy}
\end{figure}

To further highlights the differences between simpler dust evolution codes, like \texttt{two-pop-py}, and full dust coagulation approaches like \texttt{DustPy}, we present in Fig.~\ref{fig:F_peb_twopop_dustpy} the pebble flux at 1 au for a disc with a $100 \mathrm{~M_\oplus}$ planet at $10 \mathrm{~au}$ for both codes. We focus here on fragile grains ($v_\mathrm{frag} = 1 \mathrm{~m~s^{-1}}$), corresponding to the left-column in the upper panel in Fig.~\ref{fig:twopop_vs_DustPy}. We clearly see the discontinuity in the pebble flux modelled by \texttt{two-pop-py}: slightly increasing the gap depth (i.e., here lowering the viscosity) halt the drift motion of the whole dust population, as it moves at a common mass-weighted average speed. Then, the only way for dust to leak towards the inner disc is with diffusion, which barely varies as simulations shown here have a constant $\delta_\mathrm{turb} = 10^{-4}$. This discontinuity in dust transport is at the origin of the sharp transition between Regime II and III discussed in \citet{mah2024mind}. Meanwhile, the pebble flux modelled by \texttt{DustPy} is consistently higher, and better captures the gradual decrease in leakiness as the dust trap gets stronger. \texttt{DustPy} thus predicts a longer-lived oxygen-rich inner disc composition for discs with dust traps \citep[i.e., Regime II extended over a much larger parameter space, see][]{mah2024mind}, as compared with predictions from \texttt{two-pop-py}.

\end{appendix}

\end{document}